\documentclass[final,1p,times]{elsarticle}

\usepackage{amsmath}

\usepackage{siunitx}
\usepackage{soul}
\usepackage{adjustbox}
\usepackage{lipsum,graphicx,float}
\usepackage{subfig,caption}
\usepackage{multirow}
\usepackage{pdfpages}
\usepackage{notoccite}
\DeclareSIUnit\atm{atm}

\usepackage{xcolor}

\journal{Elsevier}

\begin{document}

\begin{frontmatter}



\title{Changes in boiling controlled by molar concentration-dependent diffusion of surfactants}


\author[inst1]{Mario R. Mata}
\affiliation[inst1]{organization={Department of Mechanical Engineering, University of Nevada, Las Vegas},
            addressline={4505 S Maryland Pkwy}, 
            city={Las Vegas},
            postcode={89154}, 
            state={Nevada},
            country={USA}}
            
\author[inst2]{Matic Može}
\affiliation[inst2]{organization={Faculty of Mechanical Engineering, University of Ljubljana},
            addressline={Aškerčeva cesta 6}, 
            city={1000 Ljubljana},
            country={Slovenia}}

\author[inst2]{Armin Hadžić}

\author[inst1]{Giseop Lee}
            
\author[inst1]{Blake Naccarato}

\author[inst1]{Isaac Berk}

\author[inst2]{Iztok Golobič}

\author[inst1]{H. Jeremy Cho}

\begin{abstract}
Boiling is a prevalent phase-change process that plays a vital role in facilitating efficient heat transfer from a heating surface. While this heat transfer mechanism is generally effective, a rapid increase in surface temperature can lead to hydrodynamic instabilities, resulting in a boiling crisis. Previous studies have shown that surfactants often improve boiling performance and change the boiling crisis behavior. Conventional wisdom in this field attributes that these changes in boiling behavior are tied to the critical micelle concentration (CMC) of the particular surfactant. However, our work reveals that these changes in boiling behavior are independent of the CMC for three nonionic surfactants across a wide range of molar concentrations. In addition, visual snapshots of the bubbling behavior indicate changes in bubble formation, such as bubble size and nucleation site density, influenced by the molar concentration-dependent diffusion timescale of surfactants. Hence, these findings offer compelling evidence that boiling behavior, encompassing both boiling performance and boiling crisis, is governed by the dynamic adsorption of surfactants rather than dictated by the CMC. This becomes evident when quantifying the heat transfer coefficient (HTC) and critical heat flux (CHF) using the logarithm of molar concentration, as predicted by theory. Building upon these findings, we propose insights for controlling when CHF modification occurs in specific scenarios involving any surfactants. These insights hold significant potential for optimizing heat transfer processes and leveraging surfactants in energy-related applications to maximize boiling efficiency.
\end{abstract}

\begin{graphicalabstract}
\includegraphics[width=\textwidth]{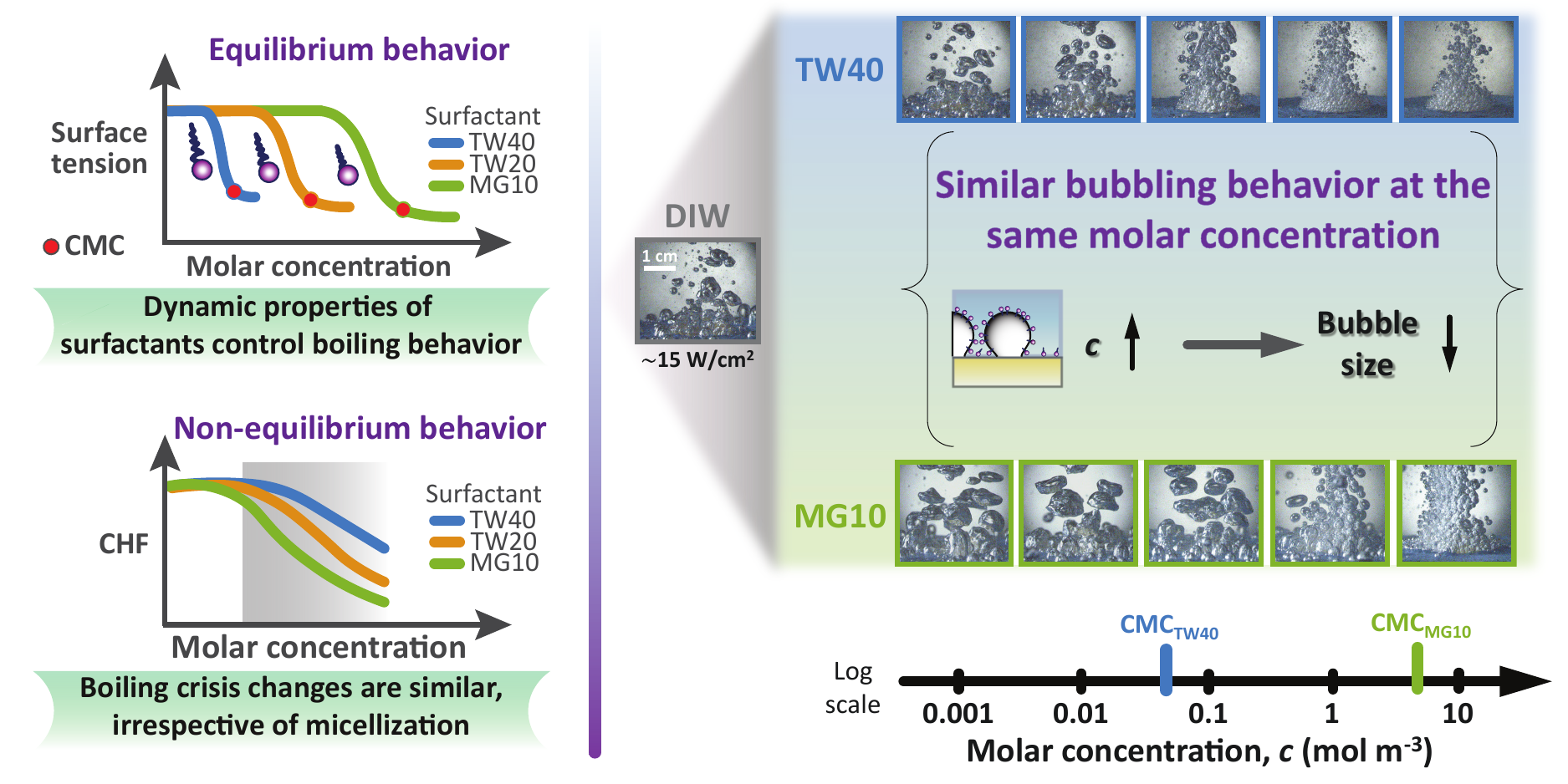}
\end{graphicalabstract}

\begin{highlights}
\item Boiling is influenced by surfactant adsorption dynamics.

\item Similar bubbling observed for different surfactants at the same molar concentrations.

\item Molar concentration-dependent diffusion governs boiling behavior.

\item Boiling trends validated with different boiler and surface type.

\end{highlights}

\begin{keyword}
Pool boiling \sep Critical heat flux \sep Boiling crisis \sep Surfactants \sep Critical micelle concentration \sep Dynamic adsorption \sep Diffusion timescale
\PACS 0000 \sep 1111
\MSC 0000 \sep 1111
\end{keyword}

\end{frontmatter}


\section{Introduction}
\label{Introduction}
Numerous industrial processes require precise temperature control for power generation and cooling systems, aimed at managing excess heat generated by appliances and equipment. Achieving such control often relies on the use of an efficient heat transfer mechanism—boiling. This versatile and vital process applies to a wide spectrum of industries and domestic applications~\cite{Beer2007, Kandlikar2004, Kanbur2020}. As heat-generating systems transfer energy to heating surfaces, localized vapor bubbles form and migrate, facilitating the extraction of heat from the surface and its subsequent dispersion into the surrounding liquid. Given the rapid advancement of technology and the increasing power consumption, leveraging the high latent heat of boiling can increase the efficiency of heat transfer in thermal systems and improve the sustainability of related technologies. To address this need, passive enhancement techniques for improving boiling performance, quantified by the heat transfer coefficient (HTC), are particularly appealing due to their relative simplicity, which can be achieved through surface modifications or the use of chemical additives. Modifying boiling surfaces alters wettability, which subsequently affects bubble behavior and heat transfer performance \cite{Bai2021, Allred2022, Liu2023}. These surface modifications include but are not limited to surfaces with micro-/nanostructures \cite{Cho2016, Liu2022, Song2022, Bongarala2022, Moze2022, Zupancic2024} and microporous materials \cite{Sarangi2016, Pham2020}---surfaces that tend to promote bubble nucleation site density, control bubble departure size, and increase bubble frequency. The impact of surface modifications on critical heat flux (CHF) varies considerably depending on the specific modification employed. CHF represents a phenomenon where the maximum heat flux in nucleate pool boiling occurs before a significant temperature increase, caused by the formation of an insulating vapor film on the boiling surface, leading to a boiling crisis. However, highly modified surfaces add significant cost and maintenance requirements \cite{Chen2020, Tang2021}; thus, chemical additives are an attractive and passive way to modify boiling behavior at large scales.

In previous studies, researchers have found that adding water-soluble polymeric or surfactant additives can modify both the HTC and CHF \cite{Cheng2007, Cho2015, Wen2022, Upadhyay2023}. Surfactants are a widely used class of molecules~\cite{Bognolo1999, Wang2021, Das2022} that adsorb to liquid-vapor \cite{Zhang2005} and solid-liquid interfaces \cite{Cho2015}, altering boiling behavior. However, there are contradictions in the reported studies. In most cases, surfactants increase the HTC when added to boiling water \cite{Wen2022, Zhang2005,  Hetsroni2009, Wasekar2002, Raza2018}. Conversely, in other instances, such as demonstrated by Xu et al. \cite{Xu2022}, the HTC decreases when using surfactants. A similar contradiction also persists with CHF. In most cases, the CHF decreases with surfactants \cite{Wen2022, Raza2018, Kang2020}. Conversely, recent work by Upadhyay et al. \cite{Upadhyay2023} shows that certain ionic-liquid surfactants can indeed increase CHF. Regardless, in all these studies, the surfactant concentration is highly fragmented, thus it is difficult to comprehend the complete dependence of HTC and CHF on concentration. This leads to an open question of how surfactant-modified boiling behavior depends on concentration. Answering this question has enormous practical utility as it will allow for the determination of optimal amounts of surfactant additives to effect the desired HTC and CHF. 

\begin{figure*}[ht]
    \centering
    \includegraphics[width=0.9\textwidth]{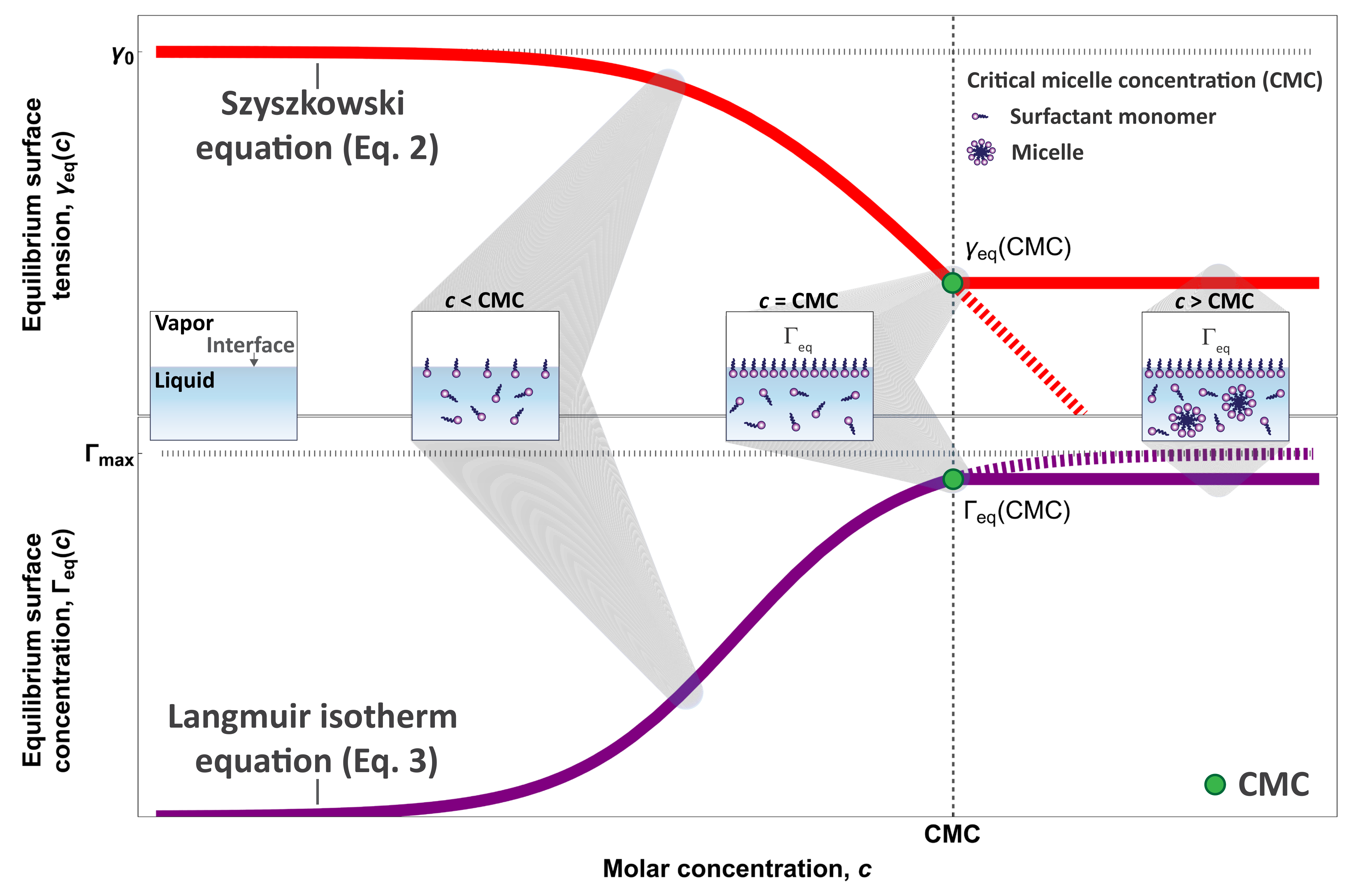}
    \caption{Surfactant monomers aggregate and form micelles once the molar concentration surpasses the CMC of a specific surfactant. Below the CMC, as concentration increases, the equilibrium surface tension, $\gamma_\text{eq}$ (Eq.~\ref{eqn:szyszkowski}), decreases and the equilibrium surface concentration, $\Gamma_\text{eq}$ (Eq.~\ref{eqn:langmuir}), increases, asymptotically approaching a theoretical $\Gamma_\text{max}$. However, above the CMC, both $\gamma_\text{eq}$ and $\Gamma_\text{eq}$ divert from Eqs.~\ref{eqn:szyszkowski} and \ref{eqn:langmuir} (dashed lines) as very few additional surfactants adsorb to the interface. Here, we depict the liquid-vapor interface; however, we expect similar behavior at the solid-liquid interface.}    
    \label{CMC}
\end{figure*}

According to previous studies, optimal boiling performance is achieved when the surfactant concentration coincides with its critical micelle concentration (CMC) \cite{Wen2022, Upadhyay2023, Zhang2005, Wasekar2002, Raza2018, Xing2020}, defining a ``conventional wisdom" for the field. The CMC is the concentration above which surfactants will aggregate into micelles (Fig.~\ref{CMC}), representing a characteristic thermodynamic property of the specific surfactant, which can vary over many orders of magnitude in concentration. However, these conventional-wisdom studies that suggest that the CMC is the optimal concentration are based on tests conducted within limited concentration ranges often near the CMC. 

With a limited concentration range tested, it is difficult to discern whether the observed optimal concentration is the true global optimum versus a local optimum. Moreover, the majority of surfactant-boiling studies present results in terms of weight parts per million (ppm or wppm). This preference is likely due to the practical convenience of ppm; however, the limitation of using ppm as the primary unit for surfactant concentration arises from the unfair comparison it creates when evaluating surfactants with different molecular weights. In the context of discussing the optimal surfactant concentration for enhancing boiling performance, using molar concentration is more appropriate (see Appendix A, Fig.~\ref{MolarMassConcentration} ).

To provide a more complete picture of the dependence of surfactant concentration on HTC, in our previous work \cite{Mata2022}, we tested four decades of concentrations, for five different surfactants, representing the most extensive mutual concentration range over the largest set of surfactant types. In contradiction to conventional wisdom, we observed that across a wide range of surfactant types with vastly different CMCs and equilibrium behaviors, HTCs started to increase around $\sim\SI{0.1}{\mole\per\meter\cubed}$ with a peak optimal around $\sim\SIrange{1}{10}{\mole\per\meter\cubed}$ \cite{Mata2022}. In essence, the concentration range optimal boiling performance can be substantially different from the CMC by several orders of magnitude. While previous work \cite{ Wen2022, Upadhyay2023,  Zhang2005, Wasekar2002, Raza2018, Xing2020} had assumed that the optimal HTC occurred around the CMC, our work \cite{Mata2022} indicates that some phenomenon other than the CMC, an equilibrium property, must be at play. We found that the timescale of dynamic adsorption of surfactants to interfaces, a non-equilibrium phenomenon~\cite{Liu2006}, is important. As with any diffusion problem, the characteristic diffusion timescale is $h^2/D$ where $h$ is the characteristic lengthscale and $D$ is the diffusion coefficient. Within the context of our surfactant diffusion problem, the characteristic lengthscale is $h=\Gamma_\text{eq}/c$~\cite{Ferri2000} where $\Gamma_\text{eq}$ is the equilibrium molar surface concentration and $c$ is the bulk molar concentration. Therefore, the surfactant diffusion timescale is 
\begin{equation}
t_\text{diff}=\frac{{\Gamma_\text{eq}}^2}{Dc^2}\text{.} \label{eqn:tdiff}
\end{equation}
However, for most surfactants, both $\Gamma_\text{eq}$ ($\sim\SI{e-6}{\mole\per\meter\squared}$) and $D$ ($\sim\SI{e-10}{\meter\squared\per\second}$) do not vary within an order of magnitude for typically applied concentration ranges. Thus, the diffusion timescale is primarily determined by the bulk molar concentration, which can vary by many orders of magnitude. In fact, $t_\text{diff}$ is strongly dependent on $c$ due to the inverse-square dependency such that a doubling of concentration results in a quadruple-fold increase in diffusion rate. When the concentration approaches around $\sim\SI{1}{\mole\per\meter\cubed}$, $t_\text{diff}$ becomes comparable to typical bubble lifetimes in pool boiling, $t_\text{b}$ ($\sim\SI{10}{\milli\second}$). When these timescales are comparable, we expect that a significant amount of surfactants adsorb to the solid-liquid and liquid-vapor interfaces that are repeatedly created and destroyed within the limited time window of $t_\text{b}$. Conversely, at low concentrations below approximately $\SI{0.1}{\mole\per\meter\cubed}$, we anticipate slow diffusion, with $t_\text{b}\ll t_\text{diff}$, resulting in minimal surfactants available for adsorption to interfaces within the limited time window of $t_\text{b}$. When a significant amount of surfactants do adsorb to interfaces ($t_\text{b}\sim t_\text{diff}$), we expect to observe changes in bubble behavior such as nucleation site density, bubble size, and non-coalescence. These adsorption-induced changes in bubble behavior ultimately determine changes in boiling performance, which we previously quantified through the HTC~\cite{Mata2022}.

While our previous work \cite{Mata2022} provides insights into how much surfactant must be added to achieve optimal HTC, our work was limited to low heat flux ranges ($<\SI{50}{\watt\per\centi\meter\squared}$) and did not test CHF. For the current study, we hypothesize that the same adsorption-induced changes in bubble behavior that alter HTC must also affect the CHF. Therefore, we expect changes in HTC and CHF to be occurring within the same concentration range for a variety of surfactants---independent of CMC---as it approaches $\sim\SI{1}{\mole\per\meter\cubed}$. This is because changes in boiling are dictated by non-equilibrium, dynamic adsorption where $\sim\SI{1}{\mole\per\meter\cubed}$ corresponds to characteristic diffusion times of surfactants migrating to the interface, $t_\text{diff}$, being similar to the timescale of the bubble cycling time in boiling, $t_\text{b}$.

\section{Boiling experiments and modeling}
\label{boiling results}
To test our hypothesis that the boiling crisis is controlled by non-equilibrium adsorption behavior, we performed flat-surface pool boiling experiments extending all the way up to the CHF with three nonionic surfactants—TWEEN-40 (TW40), TWEEN-20 (TW20), and MEGA-10 (MG10)—over a wide range of concentrations (from $\SIrange{0.001}{1}{\mole\per\meter\cubed}$), spanning four orders of magnitude. We used two different boiling devices located in separate laboratories (Boiler 1: see Appendix A, Fig.~\ref{BoilerSchematic} and Fig.~\ref{BoilerEquipment}; Boiler 2: see Appendix B, Fig.~\ref{MaticBoiler}), each featuring a flat boiling surface as it is more informative for a wide range of boiling systems compared to the cylindrical surface we used in our previous study \cite{Mata2022}. 

\begin{figure*}[ht]
    \centering
    \includegraphics[width=0.9\textwidth]{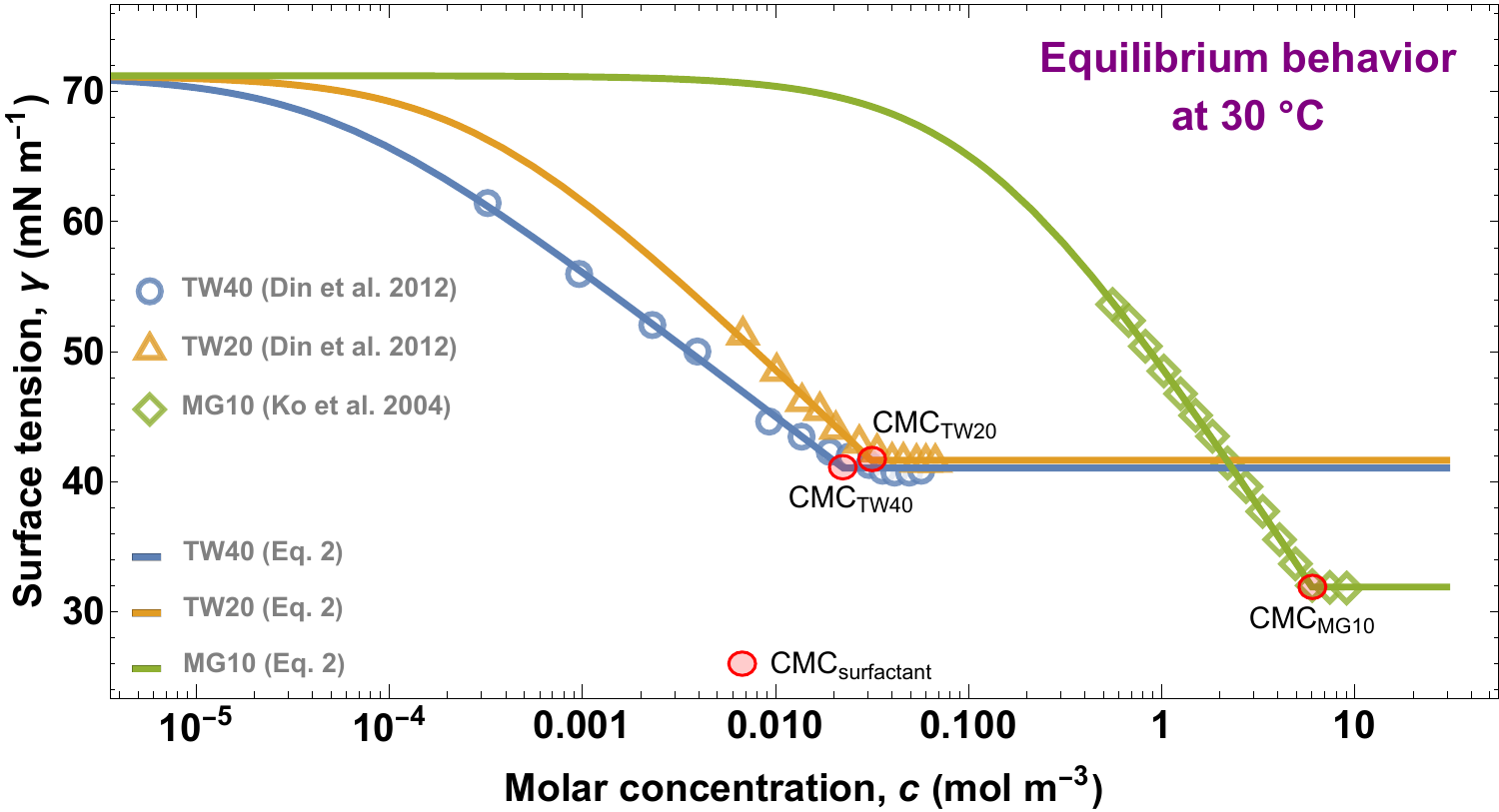}
    \caption{The equilibrium surface tension of aqueous solutions illustrates the equilibrium behavior of surfactants as we increase the molar concentration. Changes in surface tension are primarily molar-concentration-dependent and only slightly temperature-dependent, as previously modeled by Cho et al.~\cite{Cho2018}. Using Eq.~\ref{eqn:szyszkowski}, we derived the solid curves by fitting experimental surface tension data, TW40 (open circles) and TW20 (open triangles), from Din et al.~\cite{Din2012}, while MG10 (open rhombi) data points are from Ko et al.~\cite{Ko2004}. After the CMC, we assumed surface tension is constant. This figure allows us to extract information about the physicochemical and dynamic adsorption properties of these surfactants at $\SI{30}{\celsius}$, as presented in Table~\ref{TAB1}.}    
    \label{SurfaceTension}
\end{figure*}

We chose these three nonionic surfactants as they are highly differentiated by their equilibrium surface tension, $\gamma$, and CMC. That is, these surfactants exhibit significantly distinct equilibrium surface tensions at equivalent molar concentrations, as illustrated in Fig.~\ref{SurfaceTension}. We limited our study to nonionic surfactants since their surface tension behavior is simpler to model with the Langmuir adsorption isotherm \cite{Cho2018}, allowing us to avoid complications due to electric field-induced adsorption that affects boiling behavior \cite{Cho2015}. To understand the differences in the equilibrium behavior of surfactants, we evaluated equilibrium surface tension using the Szyszkowski equation \cite{Szyszkowski1908}:

\begin{equation}
{\gamma_\text{eq} = \gamma_\text{0} - \textit{R} \textit{T} \Gamma_\text{max} \ln(\text{1} + \textit{K}_\text{L} \textit{c})}\text{.} \label{eqn:szyszkowski}
\end{equation}
Here, $\gamma_\text{0}$ is the surface tension of pure water, $\textit{R}$ is the ideal gas constant, $\textit{T}$ is the absolute temperature, $\Gamma_\text{max}$ is the theoretical maximum surface concentration, and $\textit{K}_\text{L}$ is the Langmuir constant that quantifies the hydrophobic/lipophilic nature of the surfactant. $K_\text{L}$, like CMC, can vary many orders of magnitude (see Table~\ref{TAB1}). The Szyszkowski equation (Eq.~\ref{eqn:szyszkowski}), quantifying the equilibrium surface tension, implies that the surfactant has an equilibrium adsorption, $\Gamma_\text{eq}$, quantified by the Langmuir isotherm \mbox{\cite{Ross1983}}:

\begin{equation}
\Gamma_\text{eq}={\Gamma_\text{max}}\frac{\textit{K}_\text{L} \textit{c}}{{1+\textit{K}_\text{L} \textit{c}}}\text{.} \label{eqn:langmuir}
\end{equation}  
Plugging in the Langmuir isotherm (Eq.\ref{eqn:langmuir}) into Eq.~\ref{eqn:tdiff}, we obtain the characteristic diffusion timescale in terms of parameters that describe the size and chemistry of the surfactant molecule ($\Gamma_\text{max}$, $K_\text{L}$, and $D$) :

\begin{equation}
\mathrm{\textit{t}_\text{diff} = \frac{\textit{h}^2}{\textit{D}} = \frac{\Gamma_\text{max}^2}{\textit{D} \textit{c}^2} \left(\frac{\textit{K}_\text{L} 
\textit{c}}{1 + \textit{K}_\text{L} 
\textit{c}}\right)^2. \label{eqn:tdiffapprox}}
\end{equation}
Using Eq.~\ref{eqn:tdiffapprox}, we determined the values of $t_\text{diff}$ for each surfactant and concentration we tested as a means to compare diffusion timescales with bubble lifetimes, $t_\text{b}$, as shown in Fig.~\ref{BoilingCurves}.

To ensure consistency in our pool boiling results, we used a flat, circular, gold boiling surface with a diameter of $\SI{2}{\centi\meter}$. Gold offers the advantage of being highly inert and less prone to corrosion compared to other materials such as copper. We characterized the surface morphology and its material composition (see Appendix A, Fig.~\ref{SEMImages}) and the wetting effects (see Appendix A, Fig.~\ref{DIWGoldCA}) that could be associated with possible surface aging and corrosion. This surface demonstrated repeatable HTC and CHF values for DI water 
averaging $5.0\pm0.1$ $\SI{}{\watt\per\centi\meter\squared\per\celsius}$ (with a variation of $\SI{2}{\percent}$) and $119.3\pm1.7$ $\SI{}{\watt\per\centi\meter\squared}$ (with a variation of $\SI{1.4}{\percent}$), respectively. These averages were computed from nine different experiments, with three experiments conducted for each DI water boiling curve, as illustrated in Fig.~\ref{BoilingCurves}.

\begin{table}[t]
\centering
  \caption{Physicochemical and dynamic adsorption properties of surfactants.}
\label{TAB1}
\begin{adjustbox}{width=\textwidth}
\begin{tabular}{cccccccc}
            \hline
            \multirow{2}{*}{\text{Surfactant}}
            & \multirow{2}{*}{\text{Abbreviation}}
            &\multicolumn{2}{c}
            {$\text{CMC}$ ($\SI{}{\mole\per\meter\cubed}$)}
            & \multirow{2}{*}{\text{Molecular weight} ($\SI{}{\gram\per\mole}$)}
            & \multirow{2}{*}{$\Gamma_\text{max}$ (10$^{-6}$ $\SI{}{\mole\per\meter\squared}$)}
            & \multirow{2}{*}{$\textit{K}_\text{L}$ ($\SI{}{\meter\cubed\per\mole}$)}
            & \multirow{2}{*}{$\textit{D}$ (10$^{-10}$ $\SI{}{\meter\squared\per\second}$)}
            \\\cline{3-4}
            & & $\SI{30}{\celsius}$ &$\SI{98}{\celsius}$ & & & & \
            \\\hline
            TWEEN-40  & TW40 & 2.25e-2 & 5.00e-2 & 1283.6 & 1.9 & 20837.5 & 2.5\\
            TWEEN-20 & TW20  & 3.14e-2 & 1.06e-1 & 1227.5 & 2.5  & 3750.4 & 2.8\\
            MEGA-10  & MG10 & 6.01e+0 & 7.37e+0 & 349.5 & 3.9 & 8.6 & 4.8\\
            \hline
            \multicolumn{8}{l}{
            \noindent
            We subtract the CMCs at $\SI{30}{\celsius}$ from the data in Fig. \ref{SurfaceTension}.}\\
            \multicolumn{7}{l}{
            To determine the CMCs at $\SI{98}{\celsius}$, we employ a predictive tool \cite{Zoeller1995}, except for TW40, where we rely on extrapolated data \cite{Mohajeri2012}.}\\ 
            \multicolumn{7}{l}{
            We determine $\Gamma_\text{max}$ and $\textit{K}_\text{L}$ by applying Eq.~\ref{eqn:szyszkowski} to the experimental data in Figure \ref{SurfaceTension}}\\
            \multicolumn{7}{l}{ 
            We use a calculation method for $\textit{D}$ \cite{Cho2018} that is suitable for various alkanes \cite{Tee1966}.
            }
\end{tabular}
\end{adjustbox}
\end{table}

\label{boiling performance and boiling crisis}

\subsection{Changes in boiling performance}

\begin{figure*}[t]
    \centering
    \includegraphics[width=0.7\textwidth]{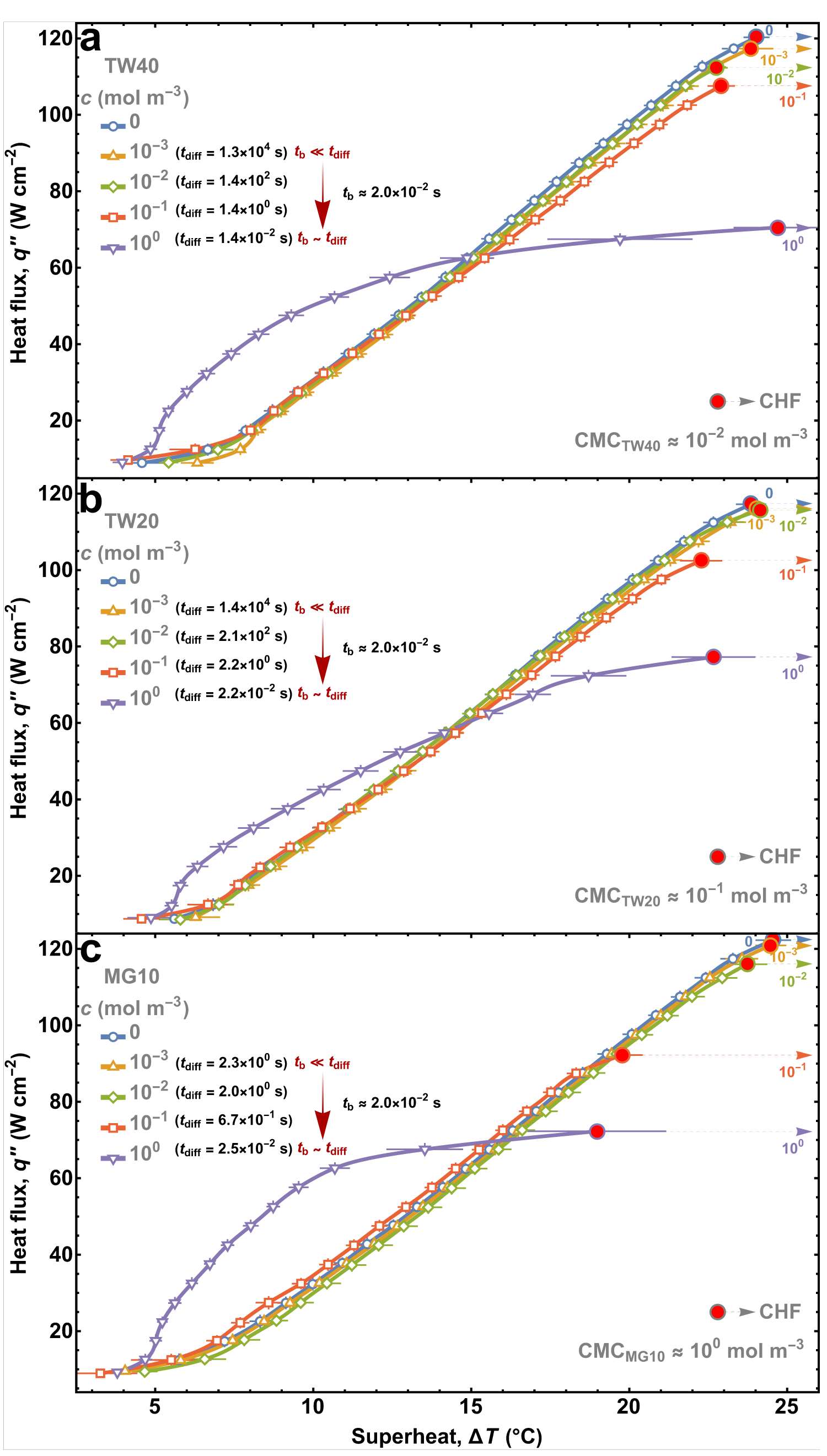}
    \caption{The boiling curves conducted on the same gold boiling surface using three distinct nonionic surfactants ((a) TW40, (b) TW20, and (c) MG10, arranged in ascending CMC order) exhibit similar concentration-dependent changes despite their vastly different CMCs. We observe significant changes in HTC and CHF at concentrations above $\sim\SI{0.1}{\mole\per\meter\cubed}$, aligning with our expectations of significant adsorption as $t_\text{diff}$ (Eq.~\ref{eqn:tdiffapprox}) approaches $t_\text{b}$. These boiling curves are averaged over three experiments per concentration. The bars accompanying the solid lines represent the uncertainty associated with each boiling curve. In Appendix A, Fig.~\ref{RohsenowFit}, we provide a comparison of the boiling curve for deionized water ($c = \SI{0}{\mole\per\meter\cubed}$) with the Rohsenow correlation.}
    \label{BoilingCurves}
\end{figure*}

We show the pool boiling curves for TW40, TW20, and MG10 in Fig.~\ref{BoilingCurves}, where $\textit{q}''$ is the heat flux and $\Delta\textit{T}$ is the superheat defined as the temperature difference between the boiling surface, $T_\text{s}$, and the saturated bulk liquid, $T_\text{sat}$. We quantified the HTC of these three different surfactants as $\textit{q}''/\Delta\textit{T}$ (for the overall HTCs, see Appendix A, Fig.~\ref{HTCPlots}). With an increase in the molar concentration from $\SI{0}{\mole\per\meter\cubed}$ to $\SI{0.001}{\mole\per\meter\cubed}$, all surfactants exhibited similar boiling behavior as quantified by slight decreases of $\SI{1.4}{\percent}$ in HTC and $\SI{1.1}{\percent}$ in CHF relative to DI water (see Fig. \ref{BoilingCurves}). This indicates that $\SI{0.001}{\mole\per\meter\cubed}$ is too low to affect boiling behavior. A small change occurred at $\SI{0.01}{\mole\per\meter\cubed}$ as quantified by decreases of $\SI{1.9}{\percent}$ in HTC and $\SI{3.6}{\percent}$ in CHF relative to DI water. An even larger change occurred when we increased the molar concentration to $\SI{0.1}{\mole\per\meter\cubed}$, as quantified by more pronounced decreases of $\SI{6}{\percent}$ in HTC and $\SI{15}{\percent}$ in CHF relative to DI water. The most substantial decreases took place when we further increased the concentration by an order of magnitude, from $\SIrange{0.1}{1}{\mole\per\meter\cubed}$. Specifically, at a molar concentration of $\SI{1}{\mole\per\meter\cubed}$, the boiling curves shifted markedly to the left, displaying lower superheat and an approximately two-fold decrease in CHF. This is in agreement with our expectation that adsorption-induced changes in HTC and CHF should occur as $t_\text{diff}$ (Eq.~\ref{eqn:tdiffapprox}) approaches $t_\text{b}$. Despite the surfactants having vastly different CMCs (see Table \ref{TAB1}), the boiling curves of all three surfactants, at four different molar concentrations, exhibited uncannily similar shapes and dependencies on concentration. This supports the idea that the overall boiling behavior for these three different surfactants is controlled by the same molar concentration-dependent diffusion timescale, $t_\text{diff}$.
\clearpage

\clearpage

\subsection{Changes in bubbling behavior}

\begin{figure*}[ht]
    \centering
    \includegraphics[width=1\textwidth]{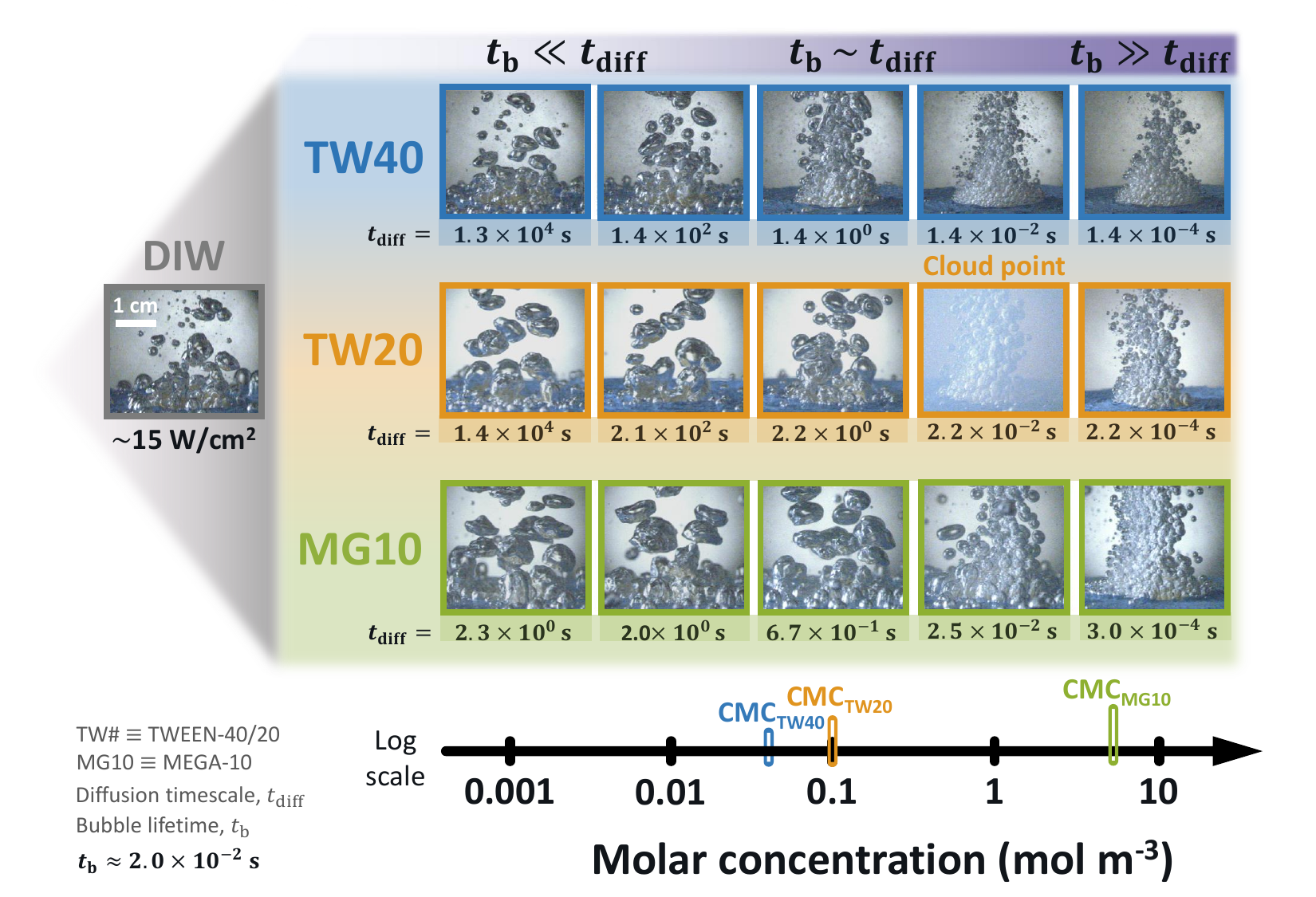}
    \caption{Snapshots of TW40, TW20, and MG10 reveal similar changes in bubbling behavior at the same molar concentrations, regardless of their CMCs. TW20 undergoes a small degree of phase separation (cloud point) at $\SI{1}{\mole\per\meter\cubed}$. At small molar concentrations, where the diffusion timescale, ${t}_\text{diff}$ (Eq.~\ref{eqn:tdiffapprox}), is slower than the bubble lifetime, ${t}_\text{b}$, changes in bubbling behavior have minimal impact on the bubble size and nucleation site density. However, as ${t}_\text{b}\sim{t}_\text{diff}$ or ${t}_\text{b}\gg{t}_\text{diff}$, a noticeable change in bubbling behavior occurs (decreased bubble size, increased nucleation site density, and increased non-coalescence), further supporting the idea that the dynamic adsorption of surfactants is dictated by the same molar concentration-dependent diffusion timescale.}
    \label{BubblingBehavior}
\end{figure*}

In further support of the idea that surfactant concentration-dependent dynamic adsorption controls boiling behavior, we observed changes in bubble nucleation site density, bubble size, and bubble non-coalescence (see Fig.~\ref{BubblingBehavior}), a similar behavior previously noted in literature \cite{Raza2018, Hetsroni2006}. We carefully selected the concentration and surfactant to minimize phase separation and cloud-point effects~\cite{AlSabagh2011}, preventing any visual blurriness. These cloud-point effects are commonly observed for nonionic surfactants, which are the same type of surfactants used in this study. Moreover, the cloud point of TW20 always consistently occurred at $\SI{1}{\mole\per\meter\cubed}$ and cleared out as soon as we increased the molar concentration of the boiling solution. Both boilers, Boiler 1 (in the US) and Boiler 2 (in Slovenia) observed the same cloud-point effects with TW20 at $\SI{1}{\mole\per\meter\cubed}$. We took images at $\SI{15}{\watt\per\centi\meter\squared}$ as this is above the onset of nucleate boiling, but a bubbling rate conducive to visual observations. We also obtained images at higher heat fluxes close to CHF (Appendix A, Fig.~\ref{CHFBubblingBehavior}), but we were not able to discern bubble behavior due to the chaotic flows and high vapor qualities that occlude the boiling surface and individual bubbles. The low-heat-flux snapshots (Fig.~\ref{BubblingBehavior}) reveal that when ${t}_\text{diff}$ is slow (quantified by Eq.~\ref{eqn:tdiffapprox}), bubble size and nucleation site density did not appear to change substantially within a molar concentration range of $\SIrange{0.001}{0.01}{\mole\per\meter\cubed}$. As we increased the molar concentration from $\SIrange{0.01}{0.1}{\mole\per\meter\cubed}$, we observed that bubble size decreased very slightly. This indicates that when ${t}_\text{b}\ll{t}_\text{diff}$, the molar concentration is too low to affect bubbling behavior. A considerable change occurred when we increased the molar concentration from $\SIrange{0.1}{1}{\mole\per\meter\cubed}$: bubble size dramatically decreased, nucleation site density increased, and bubble non-coalescence increased. This starkly different bubbling behavior emerged as ${t}_\text{diff}$ became faster (Eq.~\ref{eqn:tdiffapprox}), approaching ${t}_\text{b}$. At an even faster ${t}_\text{diff}$, increasing the molar concentration from $\SIrange{1}{10}{\mole\per\meter\cubed}$, resulted in diffusion timescales faster than the bubble lifetime, ${t}_\text{b}\gg{t}_\text{diff}$, where we observed similarly dramatic decreases in bubble size, increases in nucleation site density, and increases in non-coalescence as compared to DI water. All these changes in bubbling behavior occurred at concentrations that are independent of the CMCs (Figs.~\ref{BoilingCurves} and \ref{BubblingBehavior}), challenging conventional wisdom and providing further evidence of the importance of the non-equilibrium dynamic adsorption of surfactants. 

\clearpage

\subsection{Direct modeling of dynamic adsorption}
\label{theory} 

To provide additional evidence that boiling behavior is controlled by non-equilibrium, dynamic adsorption, we directly modeled this transient adsorption process to obtain surface concentration, $\Gamma(t)$, as a function of time. We did this by numerically solving the Ward-Tordai equation~\cite{Ward1946} where
\begin{subequations}
\begin{equation}
\Gamma(\textit{t})=2\textit{c}\sqrt{\frac{\textit{D}\textit{t}}{\pi}}-\sqrt{\frac{\textit{D}}{\pi}}\int_{0}^{t} \frac{c_0(\tau)}{\sqrt{\textit{t}-\tau}} \,d\tau
\text{.} \label{eqn:wardtordai}
\end{equation} 
Here, $\tau$ is an integration dummy variable. The Ward-Tordai equation is the exact solution of the transient, one-dimensional diffusion problem of surfactants migrating toward an interface created at $t=0$. Using the Langmuir isotherm (see Eq.~\ref{eqn:langmuir}) we can express the subsurface concentration, $c_0(\tau)$, as
\begin{equation}
c_0(\tau)=\frac{\Gamma(\tau)/\Gamma_\text{max}}{\textit{K}_\text{L}(1-\Gamma(\tau)/\Gamma_\text{max})} 
\text{} \label{eqn:dummylangmuir}
\end{equation}
assuming local equilibrium between the interface and the subsurface. By nondimensionalizing dynamic adsorption as $\Gamma(t)/\Gamma_\text{max}$, subsurface concentration as $c_0/c$, and time (interfacial age) as $tDc^2/\Gamma_\text{max}^2$, we can nondimensionalized the entire Ward-Tordai equation (Eq.~\ref{eqn:wardtordai}), with the Langmuir isotherm (Eq.~\ref{eqn:langmuir}) incorporated, as

\begin{equation}
\frac{\Gamma(T)}{\Gamma_\text{max}} = 2\sqrt{\frac{T}{\pi}}-\frac{1}{\sqrt\pi}\int_{0}^{T} \frac{\Gamma(T')/\Gamma_\text{max}}{\textit{K}_\text{L}c(1-\Gamma(T')/\Gamma_\text{max})\sqrt{{T}-{T'}}} \,d{T'}
\text{.} \label{eqn:wardtordaimodified}
\end{equation}
\end{subequations}
Here, $T$ is a dimensionless time where $T \equiv tDc^2/\Gamma_\text{max}^2$ and $T'$ is its associated dummy variable. We used a numerical method to perform the integration of Eq.~\ref{eqn:wardtordaimodified} using an algorithm originally developed by Li et al.~\cite{Li2010} (see Appendix A, Fig.\ref{Code1} for our Mathematica code). By integrating Eq.~\ref{eqn:wardtordaimodified} over successive times with an initial condition of $\Gamma(0)=0$, we can obtain $\Gamma(t)$ as a function of time. The independent variables include the selection of surfactant, which determines the value of $K_\text{L}$, $D$ and $\Gamma_\text{max}$ (see Table~\ref{TAB1}), and the bulk molar concentration, $c$. The bubble age, denoted as $t$, represents the time elapsed since the creation of the bubble. We denoted the time to bubble departure as the bubble lifetime, $t_\text{b}$. This lifetime is estimated to be $\SI{20}{\milli\second}$ based on Cole's work~\cite{Cole1967}. 

\begin{figure*}[ht]
    \centering
    \includegraphics[width=\textwidth]{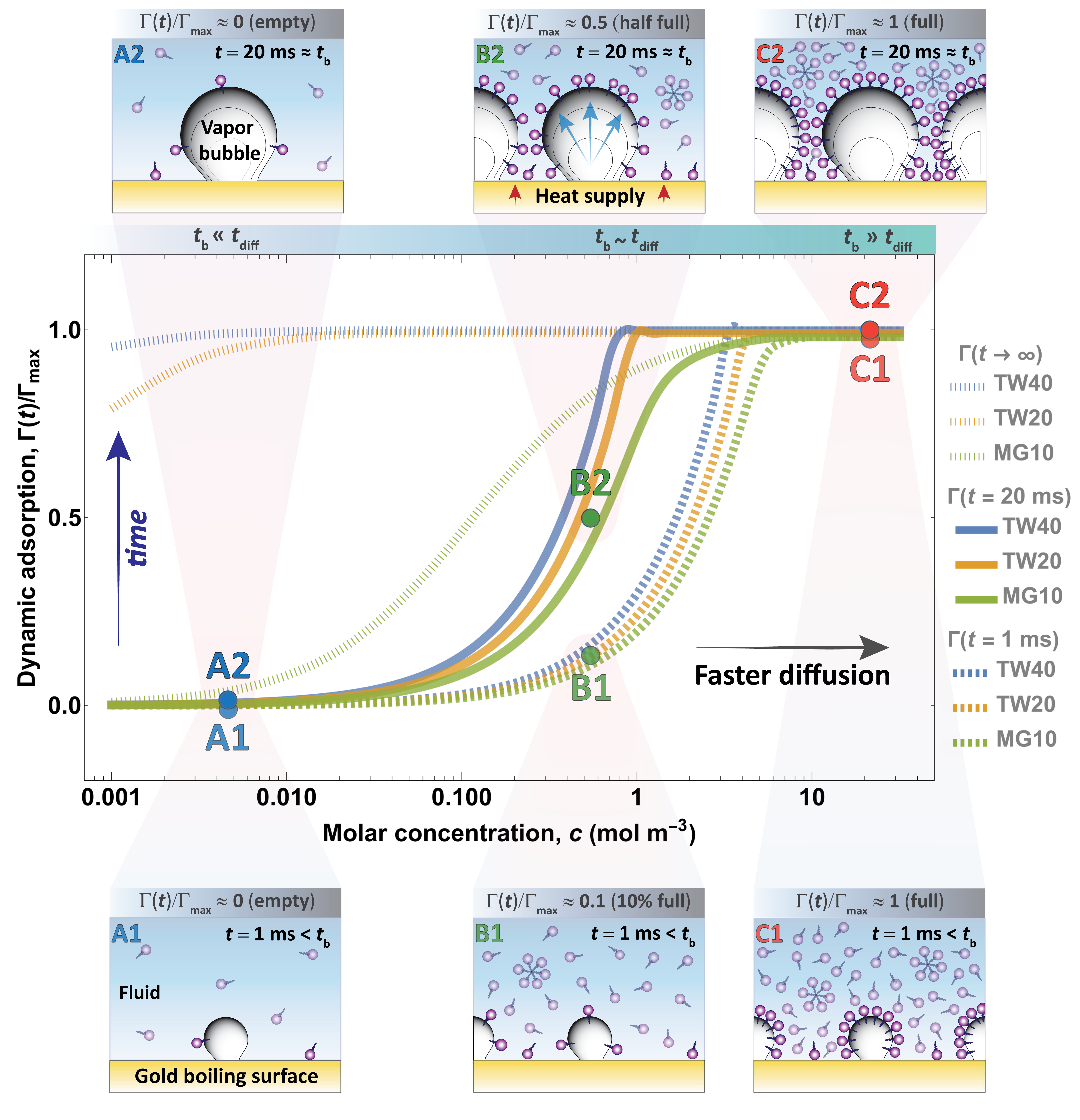}
    \caption{Changes in boiling behavior are attributed to the dynamic adsorption of surfactants to the interfaces, which exist between the solid-liquid and liquid-vapor phases. As illustrated in (A1-A2), at low molar concentration and when diffusion is slow (${t}_\text{b}\ll{t}_\text{diff}$), a few surfactants adsorb to the interfaces as the bubble grows and after the bubble lifetime, $t_\text{b}$ ($\SI{20}{\milli\second}$). However, as we increase the molar concentration in (B1-B2) and diffusion becomes fast enough that approaches $t_\text{b}$ (${t}_\text{b}\sim{t}_\text{diff}$), there is a higher interfacial concentration and the nucleation site density increases on the boiling surface. At even higher concentrations in (C1-C2), both interfacial concentration and nucleation site density increase due to faster diffusion rates (${t}_\text{b}\gg{t}_\text{diff}$). Here,  $\Gamma(t\rightarrow\infty)$ represents the equilibrium surface concentration, $\Gamma_\text{eq}$, modeled using the Langmuir isotherm (Eq.~\ref{eqn:langmuir}). All the three surfactants (TW40, TW20, and MG10), match dynamic adsorption for a concentration $\gg\SI{1}{\mole\per\meter\cubed}$. To model dynamic adsorption as the bubble grows and reaches bubble lifetime, $\Gamma(t)/\Gamma_\text{max}$, we used the Ward-Tordai equation (Eq.~\ref{eqn:wardtordaimodified}).
    }
    \label{DynamicAdsorptionPlot}
\end{figure*}

For low molar concentrations ($\sim\SIrange{0.001}{0.1}{\mole\per\meter\cubed}$), $\SI{1}{\milli\second}$ after bubble creation (Fig.~\ref{DynamicAdsorptionPlot}A1), there are few surfactants adsorbed due to slow diffusion where the diffusion time is much longer than the bubble lifetime (${t}_\text{b}\ll{t}_\text{diff}$). As the bubble grows over time and reaches the bubble lifetime, $t_\text{b}$, of $\SI{20}{\milli\second}$, adsorption is still low due to this slow diffusion as illustrated in Fig.~\ref{DynamicAdsorptionPlot}A2. For higher molar concentrations ($\sim\SIrange{0.1}{1}{\mole\per\meter\cubed}$), $\SI{1}{\milli\second}$ after bubble creation (Fig.~\ref{DynamicAdsorptionPlot}B1), there is a slight increase of surfactant adsorbed due to faster diffusion where the diffusion time is on the order of the bubble lifetime (${t}_\text{b}\sim{t}_\text{diff}$). There is an even higher interfacial concentration as the bubble age reaches the bubble lifetime, $t_\text{b}$, due to more time allowed for this fast diffusion to occur as illustrated in Fig.~\ref{DynamicAdsorptionPlot}B2. For higher molar concentrations ($\sim\SI{10}{\mole\per\meter\cubed}$), the interface is already nearly full even $\SI{1}{\milli\second}$ after bubble creation (Fig.~\ref{DynamicAdsorptionPlot}C1) because the diffusion is extremely fast where the diffusion is time is much shorter than the bubble lifetime (${t}_\text{b}\gg{t}_\text{diff}$). Since the interface is already nearly equilibrated it remains at this near-full state as the bubble ages towards its lifetime as illustrated in Fig.~\ref{DynamicAdsorptionPlot}C2. In Fig.~\ref{DynamicAdsorptionPlot}, as shown by the equilibrium ($\Gamma(t\rightarrow\infty)$; thinly hashed lines) and bubble-lifetime dynamic adsorption ($\Gamma(t=\SI{20}{\milli\second})$; solid lines) curves, dynamic adsorption matches equilibrium values at around $\sim\SI{1}{\mole\per\meter\cubed}$ for all three tested surfactants. This detailed dynamic adsorption modeling results confirm that diffusion speed is primarily dictated by molar concentration as quantified Eq.\ref{eqn:tdiff}. 

\clearpage
\subsection{Relating dynamic adsorption to boiling}
\begin{figure*}[t]
    \centering
    \includegraphics[width=0.65\textwidth]{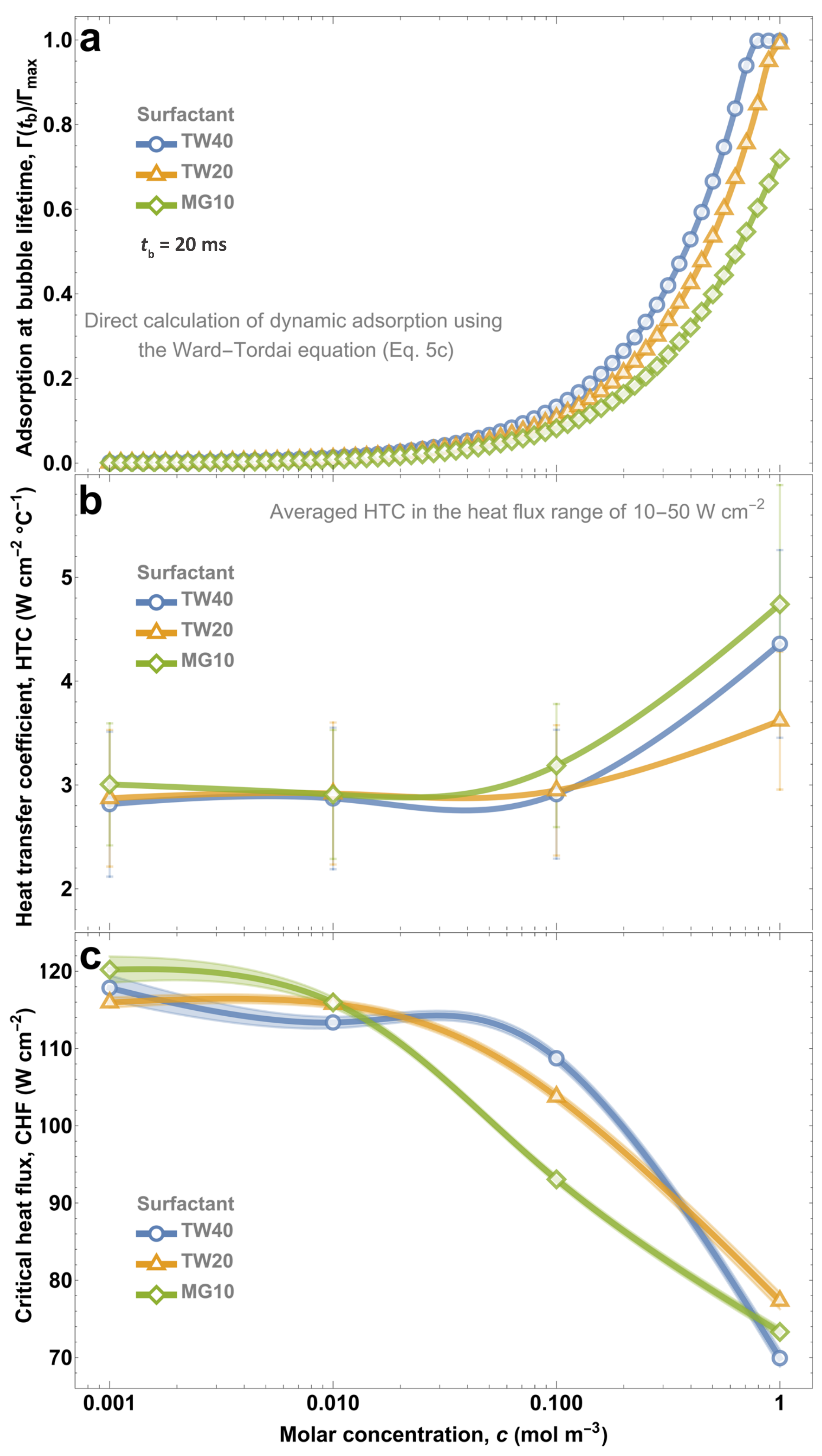}
    \caption{We use the dynamic adsorption properties of the surfactants shown in Table~\ref{TAB1} to numerically solve the Ward-Tordai equation (Eq.\ref{eqn:wardtordaimodified}). We quantify dynamic adsorption in (a) for a bubble lifetime of approximately $\sim\SI{20}{\milli\second}$. The increase in adsorption to interfaces is a direct result of an increase in the molar concentration, driven by the faster diffusion rate as defined in Eq.~\ref{eqn:tdiffapprox}. It is noteworthy that all surfactants exhibit a change in dynamic adsorption around $\sim\SI{0.1}{\mole\per\meter\cubed}$, which aligns with the changes in (b) HTC and (c) CHF, respectively. In (b), we illustrate the HTC, which exhibits a significant change from $\SIrange{0.1}{1}{\mole\per\meter\cubed}$. Similarly, in (c), we observe a significant change in CHF from $\SIrange{0.01}{0.1}{\mole\per\meter\cubed}$, which follows a similar trend across the three different surfactants. These changes in HTC and CHF are supported by our dynamic framework, which indicates that the adsorption of surfactants to interfaces governs boiling behavior and is highly dependent on the molar concentration, as demonstrated with the Ward-Tordai numerical solution in (a).}
    \label{MasterPlot}
\end{figure*}

To connect how dynamic adsorption changes correlate with boiling HTC and CHF, we show these quantities as a function of molar concentration in Fig.~\ref{MasterPlot}. Changes in HTC and CHF are small in the limit at low concentrations ($\SIrange{0.001}{0.01}{\mole\per\meter\cubed}$) This is due to slow diffusion as illustrated in Fig.~\ref{DynamicAdsorptionPlot} and as quantified in Fig.~\ref{MasterPlot}a where dynamic adsorption is low. At higher concentrations ($\SIrange{0.1}{1}{\mole\per\meter\cubed}$) where diffusion is faster, the dynamic adsorption is substantially higher. As this dynamic adsorption increases with molar concentration, the HTC (see Fig.~\ref{MasterPlot}b) increases synchronously. Similarly, changes in CHF (see Fig.~\ref{MasterPlot}c) occur synchronously with higher dynamic adsorption.

\clearpage

\begin{figure*}[ht]
    \centering   \includegraphics[width=\textwidth]{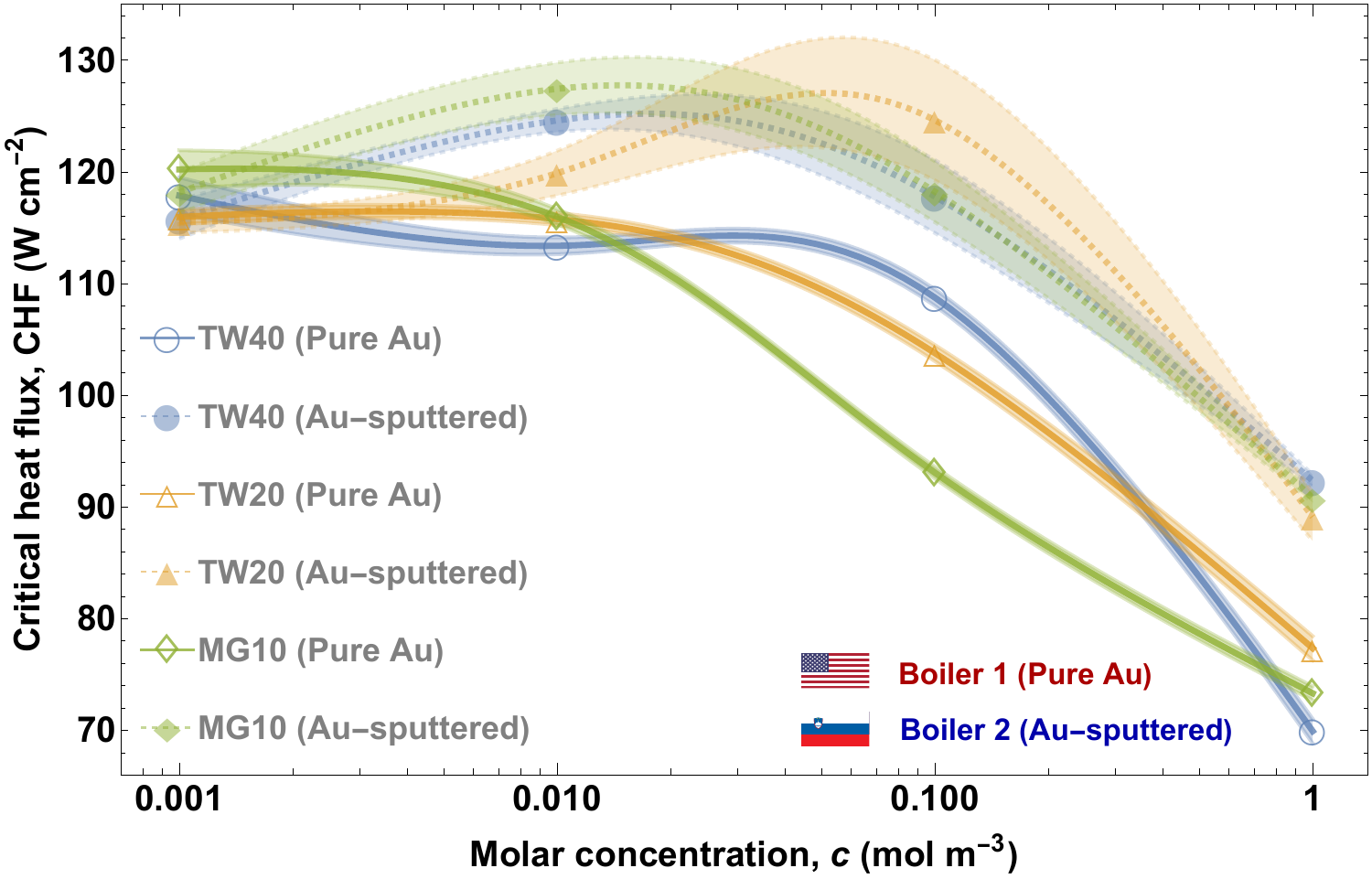}
    \caption{Validation of CHF results between two different boilers. The first boiler (Boiler 1, see Appendix A, Fig.~\ref{BoilerSchematic}) used a pure-gold boiling surface and the second boiler (Boiler 2, see Appendix B, Fig.~\ref{MaticBoiler}) used a gold-sputtered copper surface. Boiler 2 results closely align with the results of Boiler 1, despite their differences in boiling apparati. Error bands represent one standard deviation in CHF changes due to averaging of $q''$.}
    \label{DKLandMaticCHFs}
\end{figure*}

\clearpage
The results we obtained are independent of the specific implementation of the boiling test apparatus. The molar concentration dependence on HTC is consistent with our earlier study~\cite{Mata2022} that used a copper, cylindrical boiling surface as opposed to the flat gold foil used in the current study. Furthermore, we found remarkable consistency for an entirely different boiling apparatus (Boiler 2, Appendix B, Fig.~\ref{MaticBoiler}) with a gold-sputtered copper surface located in a separate laboratory at the University of Ljubljana, Slovenia. Across all tested surfactants, we observed the same synchronous changes in CHF and HTC (see Fig.~\ref{DKLandMaticCHFs} and Appendix B, Fig.~\ref{BCsMatic}) as Boiler 1 (Appendix A, Fig.~\ref{BoilerSchematic}). This further verifies that regardless of boiling surface material, geometry, boiling apparatus, and testing location, the molar concentration-dependent diffusion controls boiling behavior.

\section{Discussion}
The close correspondence between molar concentration and boiling performance provides compelling evidence of a diffusion-transport-limited mechanism controlling a set of phenomena relevant to boiling performance. Setting a molar concentration determines the speed of surfactants diffusing to interfaces, characterized by $t_\text{diff}$ (Eq.~\ref{eqn:tdiff}) and directly modeled using the Ward-Tordai equation (Eq.~\ref{eqn:wardtordaimodified}). As surfactants dynamically adsorb to interfaces, they reduce surface tension to values that could be far from equilibrium---in fact, we modeled dynamic surface tension as illustrated in Appendix A, Fig.~\ref{DynamicSurfaceTension} using our Mathematica code disclosed Fig.~\ref{Code2}. The presence of adsorbed surfactants also increases bubble non-coalescence, which we directly observed in Fig.~\ref{BubblingBehavior}. The lower surface tension and increase in non-coalescence effectively reduce the observed bubble size (Fig.~\ref{BubblingBehavior}) due to earlier bubble departure and reduced merging of bubbles. Simultaneously, the adsorption of surfactants to solid-liquid interfaces increases nucleation site density~\cite{Cho2015}, which we also observed (Fig.~\ref{BubblingBehavior}). With dramatically different bubble behavior when a substantial amount of surfactants are adsorbed, this leads to significantly different boiling performance as quantified by HTC and CHF (Fig.~\ref{MasterPlot}). In particular, we confirm that the boiling crisis is dependent on the same surfactant-transport phenomena as the HTC. Thus, the diffusion-transport-limited mechanism of boiling modification with surfactants is supported by the range of testing, modeling, and analyses performed in this work. As a result, of diffusion-transport-limited behavior, non-equilibrium, dynamic adsorption controls boiling performance. This is in direct contrast with the conventional wisdom that claims the importance of equilibrium adsorption on boiling performance as determined by CMC.  

Future studies could verify whether the diffusion-transport-limited mechanism holds true to other boiling systems. Our testing was performed at atmospheric pressure; however, sub- and super-atmospheric pressures would change the bubble lifetime. Thus, the concentration ranges where dynamic adsorption is significant would vary with the saturation pressure. For instance, for near-critical boiling applications (e.g. nuclear power plants) we expect the need for higher concentrations of surfactants to provide the necessary diffusion speeds to match the shorter bubble lifetimes. In addition, our study was limited to nonionic surfactants. Introducing ionic surfactants would enable separate control of solid-liquid and vapor-liquid adsorption through the use of electric fields~\cite{Cho2015}. Thus, bubble size and nucleation site density could presumably be separately controlled as well. Certain ionic surfactants show promise as they simultaneously increase boiling HTC and CHF at certain concentrations~\cite{Upadhyay2023}; therefore, a verification of the diffusion-transport-limited behavior could provide valuable insight. However, the use of ionic surfactants presents a significant complication as they cannot be readily modeled using the Ward-Tordai equation and the Langmuir isotherm. Alternative isotherms suitable for ionic surfactants, such as the one developed by us previously~\cite{Cho2018}, could be employed. Thus, our current work encourages a range of exciting future studies involving surfactants modifying boiling behavior. 

\section{Conclusions}
\label{Conclusions}
In this study, we investigated how surfactants can modify boiling performance. Specifically, we focused on how HTC and CHF depend on the molar concentration as this will provide practical utility in identifying optimal amounts of surfactants. To provide general insight across a range of surfactant types and boiling systems, we conducted flat-surface pool boiling experiments with three nonionic surfactants (TW40, TW20, and MG10) with highly differentiated CMCs in two completely different boiling apparati. Contrary to conventional wisdom, which suggests that changes in boiling behavior are primarily controlled by the equilibrium property of CMC, we find that the non-equilibrium, dynamic adsorption of surfactants controls boiling behavior. Substantial changes in HTC and CHF occur at concentrations around $\sim\SI{0.1}{\mole\per\meter\cubed}$ or more, delineating a regime where the timescale of diffusion is short enough to cause significant adsorption to interfaces within the limited time window of the bubble lifetime. At lower concentrations, slow diffusion causes few surfactants to adsorb, causing minimal changes in boiling performance. The results of direct modeling of diffusion using the Ward-Tordai equation confirm that concentration-dependent changes in dynamic adsorption are synchronized with the experimentally observed changes in HTC and CHF. At higher concentrations where diffusion is fast enough to cause significant adsorption, we expect changes in bubble size, nucleation site density, and non-coalescence, which we indeed observed synchronously at the same molar concentrations across all the tested surfactants. These concentration-dependent changes in bubble behavior are expected to alter boiling heat transfer, which we characterized through boiling curves. There is an uncanny resemblance in these concentration-dependent boiling curves across the three surfactants tested despite their vast differences in CMC. Our results strongly indicate that the conventional wisdom on the importance of CMC in boiling is invalid. Rather, we provide a dynamic, non-equilibrium perspective where changes in boiling are controlled by molar concentration-dependent diffusion timescales. This work offers valuable insights for applications where heat transfer can be enhanced using surfactant additives.

\section{Materials and methods}

\subsection{Boiling setup}
\label{[boiling setup]}
We conducted boiling experiments at atmospheric pressure (1 atm) using $\SI{750}{\milli\liter}$ of deionized (DI) water or aqueous solutions of surfactants within a customized boiling device (see Figs. \ref{BoilerSchematic} and \ref{BoilerEquipment}). A specially designed stainless steel coiled condenser is positioned atop one of the sealing plates in the setup. This condenser efficiently condenses incoming vapor from the experiments and ensures a constant bulk fluid level. Another crucial aspect of using a coiled condenser is to prevent changes in concentration during boiling experiments. Surfactants have a significantly higher boiling point than water due to their lower vapor pressure. Consequently, surfactants stay dissolved in the liquid phase without evaporating. In an open system where water is allowed to evaporate, the concentration of surfactants would gradually increase over time. This is because the amount of water decreases over time while the moles of surfactants remain constant. In this manner, the working fluid's vapor recondenses and returns to the liquid, ensuring a constant amount of working fluid in our boiler's closed-loop system over time.

A liquid heating bath, containing circulating ethylene glycol on the outer surfaces of the main chamber, is maintained at a higher temperature than the boiling point of the liquid under investigation. This configuration serves to preheat the liquid and uniformly sustains it under saturation conditions. There are several advantages to this configuration compared to one relying on directly heating the fluid using one or several immersed cartridge heaters. One benefit of the liquid heat bath is that heat diffuses from an outer chamber, where the bath temperature is precisely controlled, towards the inner chamber where the targeted working fluid is situated. The outer and inner chambers share the same dividing wall, ensuring uniform and even heating throughout the boiling fluid. The liquid heating bath establishes a near-perfect temperature wall boundary condition for the working fluid, resulting in nearly ideal saturation conditions with minimal heat losses. Another advantage is that bubbles, which commonly form on immersed cartridge heaters, tend to interfere with the formation of growing bubbles on the boiling surface during the analysis. Uneven temperature gradients in the working fluid, a concern with immersed cartridge heaters, are minimized by the liquid heating bath, which heats the entire wall of the inner chamber. Additionally, an immersed cartridge heater could disrupt the concentration gradients of surfactants, impacting the flow of the working fluid due to the rapid formation and oscillation of bubbles on the cartridge heaters. Nonetheless, the use of a liquid heating bath, as opposed to immersed heaters, does not alter the experimental results. This is evident in the experiments conducted on Boiler 2 in Slovenia, where a bath was not employed (see Appendix B, Figs.~\ref{MaticBoiler} and ~\ref{BCsMatic}), yet similar results were achieved compared to Boiler 1 (see Fig.~\ref{DKLandMaticCHFs}).

The boiling surface (gold foil), with a diameter of $\SI{2}{\centi\meter}$, is soldered onto a copper foil, which is then soldered to a custom-machined copper rod $\SI{9}{\centi\meter}$ long. At the base of the setup, there is a heating copper block containing eight cartridge heaters (Chromalox, CIR-3018/120 V/200 W) interconnected in a parallel circuit, collectively producing an output power of $\SI{1600}{\watt}$. This heating copper block is positioned on a custom T-slotted structure, equipped with two linear bearing mounts (80/20 part No. 6425), allowing it to be manually disconnected when observing the boiling crisis phenomenon or whenever necessary. The high purity of our gold boiling surface offers the advantage of being highly inert and less prone to corrosion compared to other materials such as copper, as we acknowledge the potential effects of oxidation \cite{YSong2022} and hydrocarbon adsorption \cite{Song2020}, which can influence boiling results.

Temperature measurements are facilitated using multiple T-type thermocouples (Omega). The thermocouples monitor the temperature within the bulk fluid, the liquid heating bath, and at various locations on the copper heating components. As for the boiling surface, four ungrounded thermocouples, each with a diameter of $\SI{1}{\milli\meter}$, are perpendicularly inserted into holes of identical dimensions located at the center of the copper rod. These thermocouples are spaced at intervals of $\SI{2.8}{\milli\meter}$ from one another (center-to-center of the thermocouple tip), with the origin situated at the boiling surface (see Appendix A, Fig.~\ref{BoilerSchematic}). Electrical signals from the thermocouples are captured by a data acquisition unit (MCC USB-3104, featuring 8 channels). Data signals are processed and precisely controlled through a Virtual Instrument (VI) utilizing LabVIEW software, which interfaces with a dedicated code we developed in Wolfram Mathematica.

\subsection{Calculation of boiling-curve data}
\label{Boiling data reduction}
The heat flux, $q''$, was calculated by linear extrapolation of the four thermocouples evenly-spaced $\SI{2.8}{\milli\meter}$ apart to the boiling surface. From these thermocouples, we obtained temperature as a function of depth into the heater, $T(x)$, using a linear regression. From this regression, we took the derivative to obtain the temperature gradient, $dT/dx$.

For a datum, $i$, we have the surface temperature of the boiling surface, ${T_{s}}_{i}$, at the position $x=0$ as follows

\begin{equation}
{T_{s}}_{i}=T(x=0) \label{eqn:surfacetemperature}
\end{equation}
and the outward heat flux for each datum is calculated as

\begin{equation}
q''_{i}=k\frac{d{T}}{d{x}} \label{eqn:heatflux}
\end{equation}
where $k$ is the thermal conductivity of the heating rod. The HTC for each datum is quantified as

\begin{equation}
\text{HTC}_{i}=\frac{q''_{i}}{{T_{s}}_{i}-T_{sat}}, \label{eqn:htc}
\end{equation}
whereas the average HTC was determined by averaging HTC values taken at different heat fluxes within the range of $\SIrange{10}{50}{\watt\per\centi\meter\squared}$ and quantified as

\begin{equation}
\overline{\text{HTC}}_{i}=\frac{\sum_{i=1}^{N}\text{HTC}_{i}}{N} \label{eqn:averagehtc}
\end{equation}
where $N$ is the number of data points. We determined the CHF as
\begin{equation}
\text{CHF}=\text{max}(q''_{i}) \label{eqn:chfmax}
\end{equation}
identifying the local maximum in the set of $q''_{i}$ data points on each boiling curve.

\subsection{Propagation of errors}
\label{Uncertainty and propagation of error calculations}
To evaluate the consistency of our experiments from one run to another, we analyzed the uncertainties across the boiling results. The error bars in Fig.~\ref{BoilingCurves} represent one standard deviation in the heat flux, $q''$, and superheat, $\Delta\textit{T}$, from averaging three experiments for each concentration. Our analysis revealed that the highest $q''$ uncertainty of $\pm\SI{1.4}{\watt\per\centi\meter\squared}$ occurred for TW20 at a molar concentration of $\SI{0.1}{\mole\per\meter\cubed}$ with a $q''$ of $\SI{102.5}{\watt\per\centi\meter\squared}$, whereas the highest $\Delta\textit{T}$ uncertainty of $\pm\SI{2.2}{\celsius}$ occured for MG10 at a molar concentration of $\SI{1}{\mole\per\meter\cubed}$ with a $\Delta\textit{T}$ of $\SI{19.0}{\celsius}$. In addition, considering the inherent nonlinearity of any boiling curve, it is expected that there will consistently be a notable error bar (see Fig.~\ref{BoilingCurves}b) associated with the averaged HTC when computed over such an extensive heat flux range that focuses on a specific plot range (non-zero starting point) to underscore variations in molar concentration.

We also assessed the repeatability of our experiments when transitioning from one surfactant to another. Before conducting a series of experiments with different surfactants at different molar concentrations, we meticulously followed a specific cleaning procedure (see~\ref{cleaningprocedures} Cleaning and testing procedures). Notably, our boiling surface did not lead to significant changes in contact angles (see Appendix A, Fig.~\ref{DIWGoldCA}). The variation in the advancing contact angle and receding contact angle before and after boiling tests was $2.2\pm\SI{0.7}{\degree}$ and $3.8\pm\SI{1.2}{\degree}$, respectively.

We conducted about 20 boiling experiments per surfactant using Boiler 1 (see CHF results in Fig.~\ref{DKLandMaticCHFs} and boiling device in Appendix A
, Fig.~\ref{BoilerSchematic}), including three sets of experiments for each concentration (taking into account reference DI water tests before and after each concentration). Similarly, complementary experiments from another boiling device, Boiler 2 (see CHF results in Fig.~\ref{DKLandMaticCHFs} and boiling device in Appendix B
, Fig.~\ref{MaticBoiler}), amounted to at least 15-20 experiments per surfactant. Combining results from both boiling devices, this amount represents a total of over 100 boiling experiments in this study.

\subsection{Sample preparation}
\label{[boiling surface]}
The boiling surface consists of a $\SI{75}{\micro\meter}$-thick $\SI{99.99}{\percent}$ gold foil (Surepure Chemetals, product No. 1506) soldered onto a $\SI{75}{\micro\meter}$-thick copper alloy 110 foil (Basiccopper). This copper foil was soldered to the top of a custom-machined long copper rod. After soldering, we carefully roughened the gold boiling surface with 320-grit (3M) sandpaper to provide nucleation spots. Then we proceeded to clean the surface with isopropyl alcohol and DI water using cotton swabs, and we dried its surface with a heat gun.

\subsection{Surfactant preparation}
\label{[stock solution]}
We purchased all three nonionic surfactants from Sigma Aldrich. Two surfactants are in liquid form, TWEEN-40 (TW40) and TWEEN-20 (TW20). The third one, MEGA-10 (MG10), comes as a powder. We dissolved all surfactants in DI water to prepare stock solutions for each surfactant, where the stock solutions of TW40 and TW20 mixtures were dissolved at a concentration of $\SI{350}{\mole\per\meter\cubed}$, and the stock solution of MG10 at a concentration of $\SI{140}{\mole\per\meter\cubed}$. We employed a precise method to dispense surfactants into the boiling solution to achieve the desired concentrations. This involved preparing a stock solution of high-concentration surfactant and micropipetting precise amounts into the solution chamber. We maintained good control over the concentration for conducting pool boiling experiments, with concentrations staying within $\pm\SI{0.4}{\percent}$. For complete dissolution, mixtures were preheated at $\SIrange{40}{50}{\celsius}$ on a hot plate and stirred with a magnetic stir bar for about 2 h. For all the experiments, we used manual micropipettes to take the right amount of the stock solution. This amount of stock solution was added to $\SI{750}{\milli\liter}$ of boiling DI water, bringing the desired molar concentration values ($\SI{0.001}{\mole\per\meter\cubed}$, $\SI{0.01}{\mole\per\meter\cubed}$, $\SI{0.1}{\mole\per\meter\cubed}$, and $\SI{1}{\mole\per\meter\cubed}$) for each tested surfactant mixture.

\subsection{Cleaning and testing procedures}
\label{cleaningprocedures}
Prior to commencing boiling experiments, we disassembled our boiler (see Fig.~\ref{BoilerSchematic}), followed by a meticulous wash using soap and water. We meticulously cleaned the gold boiling surface with isopropyl alcohol and DI water. Once we inspected the boiling surface, swapped old o-rings with new ones, re-applied the vacuum grease, connected fittings, and sealed the boiler, we added and drained $\SI{20}{\liter}$ of water inside the main tank to remove any possible surfactant residue from the wash. Next, we added $\SI{750}{\milli\liter}$ of DI water, waited until it reached boiling conditions, and recorded the respective boiling curves, in most cases, up to 10 boiling experiments one right after another. Comparing these boiling curves served as an indication that the boiler was properly cleaned.

As an important note when working with surfactants, consistency in the cleaning procedures is crucial. Once we finished testing a particular surfactant at a desired molar concentration, we would repeat the add/drain procedure for every concentration. For example, if we tested a surfactant at $\SI{0.001}{\mole\per\meter\cubed}$ and wanted to move to a higher concentration of $\SI{0.01}{\mole\per\meter\cubed}$, we would drain all the aqueous solution, then top off the boiler with approximately $\SI{2.5}{\liter}$ of water (which is the maximum volume the inside tank can hold with liquid), drain it, and repeat the process about eight times (totaling around $\sim\SI{20}{\liter}$ of added/drained water). This quantity is what we found to work effectively in our boiler in all cases, ensuring that even the highest tested concentration of $\SI{1}{\mole\per\meter\cubed}$ will be removed without leaving any surfactant residues inside the tank for subsequent tests. 

For maximum repeatability between tests to compare results, we ran experiments on the same day, at one-week intervals. For example, on a given Friday, we would test surfactant TW40 at concentrations ranging from $\SIrange{0.001}{1}{\mole\per\meter\cubed}$, drain the tested aqueous solution, and rinse the boiler with the respective $\sim\SI{20}{\liter}$ of water. The boiler would remain inactive for the following two days during the weekend. On Monday, we would conduct boiling experiments with just DI water to check its boiling performance. After another two days of inactivity, on Wednesday, we would test DI water again. Then, on the following Friday, we would carry out boiling experiments with TW20. We followed the same cleaning and testing procedures for MG10.

\subsection{Visualization}
\label{images and videos}
Bubbling events were recorded using two types of cameras. With the first camera (FLIR Blackfly S, BFS-U3-51S5), we captured a sequence of polarized images of bubbles at specific heat flux intervals for all the surfactants tested. With the second camera (Photron FASTCAM Mini AX200), we captured high-speed videos at a resolution of 512x512 and a frame rate of 10,000 frames per second. We used two LED units: one placed opposite the camera and the other positioned on the side window of the boiling device.

\subsection{Field emission scanning electron microscopy (FESEM)}
\label{FESEM}
We conducted surface characterization of our gold boiling surface with field emission scanning electron microscopy (FESEM) using a TESCAN CLARA ultra-high-resolution scanning electron microscope (see Appendix A, Fig.~\ref{SEMImages}). This microscope is equipped with an electron dispersion X-ray spectroscopy (EDS) probe to measure the chemical composition of the gold samples. Boiling performance is strongly influenced by the interaction between the surface and surfactants.

In the microscope chamber, we inserted for testing a piece of gold foil treated with the same sanding and cleaning procedures as described in the sample preparation section. The first test served to visualize a gold sample that was not submerged in boiling DI water (pre-boiling (PRE)), a second sample that was submerged in boiling water (post-boiling (POST)), a third sample that was exposed to the boiling aqueous solution of MG10 at a molar concentration of $\SI{1}{\mole\per\meter\cubed}$ (post-boiling (MG10)). 

The FESEM analysis (see Appendix A, Fig.~\ref{SEMImages}a) revealed no observable changes in the surface structure, revealing that either there was no oxidation or only minimal impact on the pure gold sample due to the exposure to boiling fluid. The EDS tests (see Appendix A, Fig.~\ref{SEMImages}b) revealed a high level of gold purity in all cases, close to $\SI{90}{\percent}$, in contrast to the labeled gold purity level of our sample of $\SI{99.99}{\percent}$ (as disclosed from Surepure Chemetals). EDS typically analyzes a very shallow layer, approximately $\SI{1}{\micro\meter}$ thick, from the surface of the sample. This high sensitivity makes it susceptible to contamination, selective erosion, and other factors that can result in chemical differences between the surface and the bulk material. Most of the remaining composition in the analysis comes from carbon adsorption on the gold samples, accounting for $\approx\SI{10}{\percent}$.

\subsection{Dynamic contact angle measurements}
\label{DCA}
We conducted dynamic contact angle measurements of our pristine gold ($\SI{99.99}{\percent}$ 24-carat purity) boiling surface with a custom in-situ goniometer (see Appendix A, Fig. \ref{DIWGoldCA}a) precisely controlled by a custom PID VI using LabVIEW and a code made in Wolfram Mathematica running in the background for live measurements. This goniometer is fixed inside the boiling setup and can be lifted up and down whenever needed to analyze the boiling surface. For the purpose of visualization, we removed parts of the boiler to clearly expose the goniometer. 

The contact angle results (see Appendix A, Fig. \ref{DIWGoldCA}b) showed that the dynamic contact angle of the gold boiling surface did not significantly change from test to test (pre/post-boiling). This ensures the reliability of scientific data and insights, emphasizing the necessity for consistent surface conditions across tests.

\section*{Conflicts of interest}
\label{Conflicts of interest}
The authors have no conflicts to declare.

\section*{Acknowledgements}
\label{Acknowledgements}
The authors acknowledge Amir Kashani for assisting in the dynamic contact angle experiments and Brandon Ortiz for ensuring proper software control of the boiler. The authors also extend their gratitude to Youngsup Song for his invaluable feedback during the early stages of this work, particularly concerning surface contamination, corrosion, and carbon adsorption effects.

Additionally, the authors would like to express their sincere appreciation for the contributions of Clark Bossert, Ryan McKenzie, and Genaro Marcial-Lorza for machining support. The authors would especially like to thank and honor the memory of the late Terry Kell, who provided invaluable wisdom, critical insight, and extremely precise machining of key components of our boiler.

This work received support from various sources, including UNLV Start-up Funds, Koshee Innovation, Nevada Gold Mines, the ACS Petroleum Research Fund Doctoral New Investigator Grant (grant No. 66043-DNI5), and the NSF CAREER Award (grant No. 2239416). The authors, M. Može, A.H, and I.G. acknowledge financial support from the state budget of the Slovenian Research and Innovation Agency (program No. P2-0223).

\appendix

\section{Additional results and experimental equipment of Boiler 1}
\label{APPENDIXA}
Here, we include supplementary data associated with the first boiler that is configured with a pure gold boiling surface. Experiments were conducted in Las Vegas, USA.

\clearpage
\begin{figure*}[t]
    \centering
    \includegraphics[width=\textwidth]{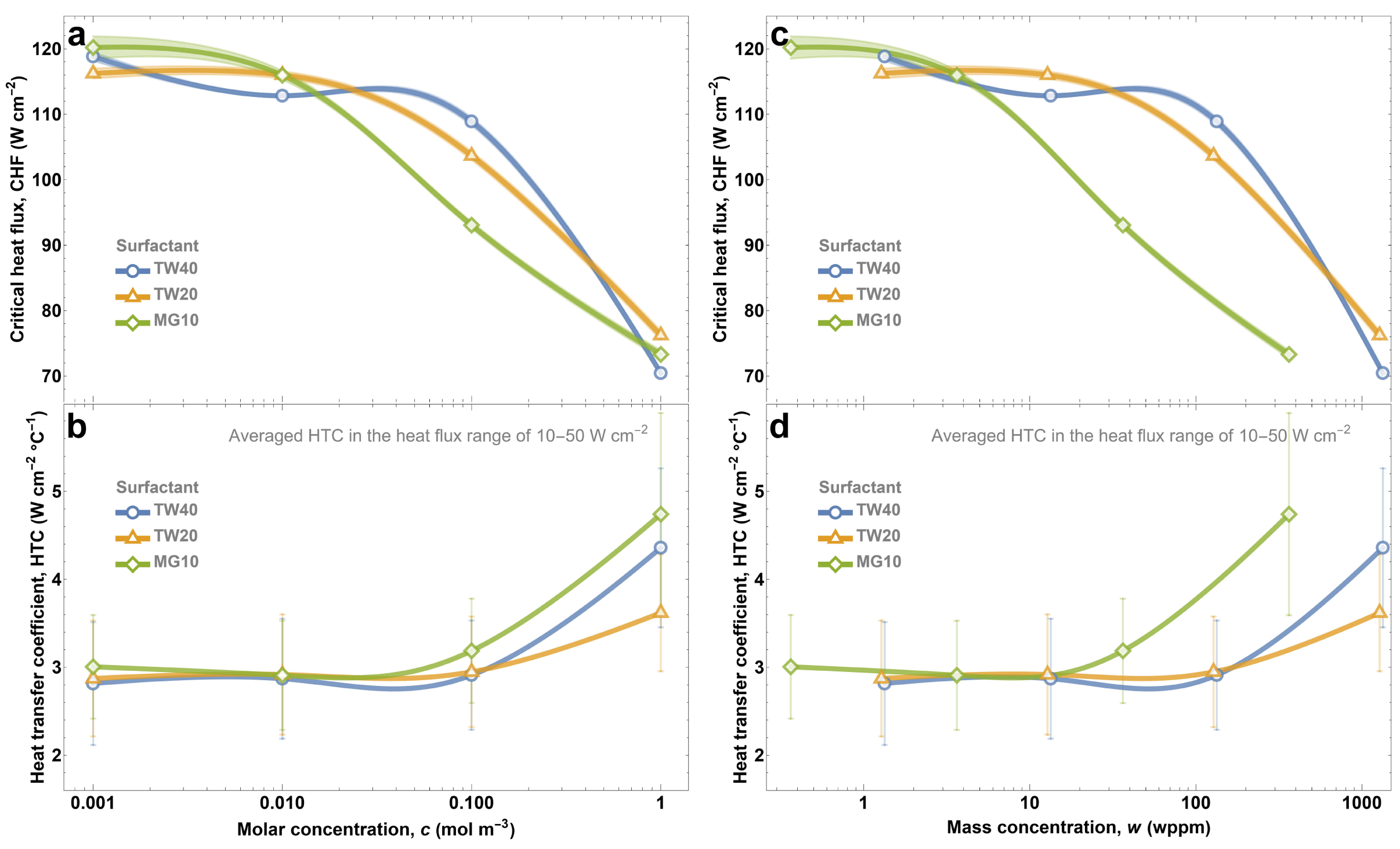}
    \caption{Comparison of the CHF and HTC versus surfactant concentration, in terms of molarity (a-b) and parts per million by weight (c-d). Note how (a-b) the start and end points of the three surfactants in the molar concentration are precisely the same. However, the same cannot be said if plotted in mass concentration, where each surfactant exhibits different start and end points, as expected for surfactants of different molecular weights.}
    \label{MolarMassConcentration}
\end{figure*}

\clearpage
\begin{figure*}[t]
    \centering
    \includegraphics[width=\textwidth]{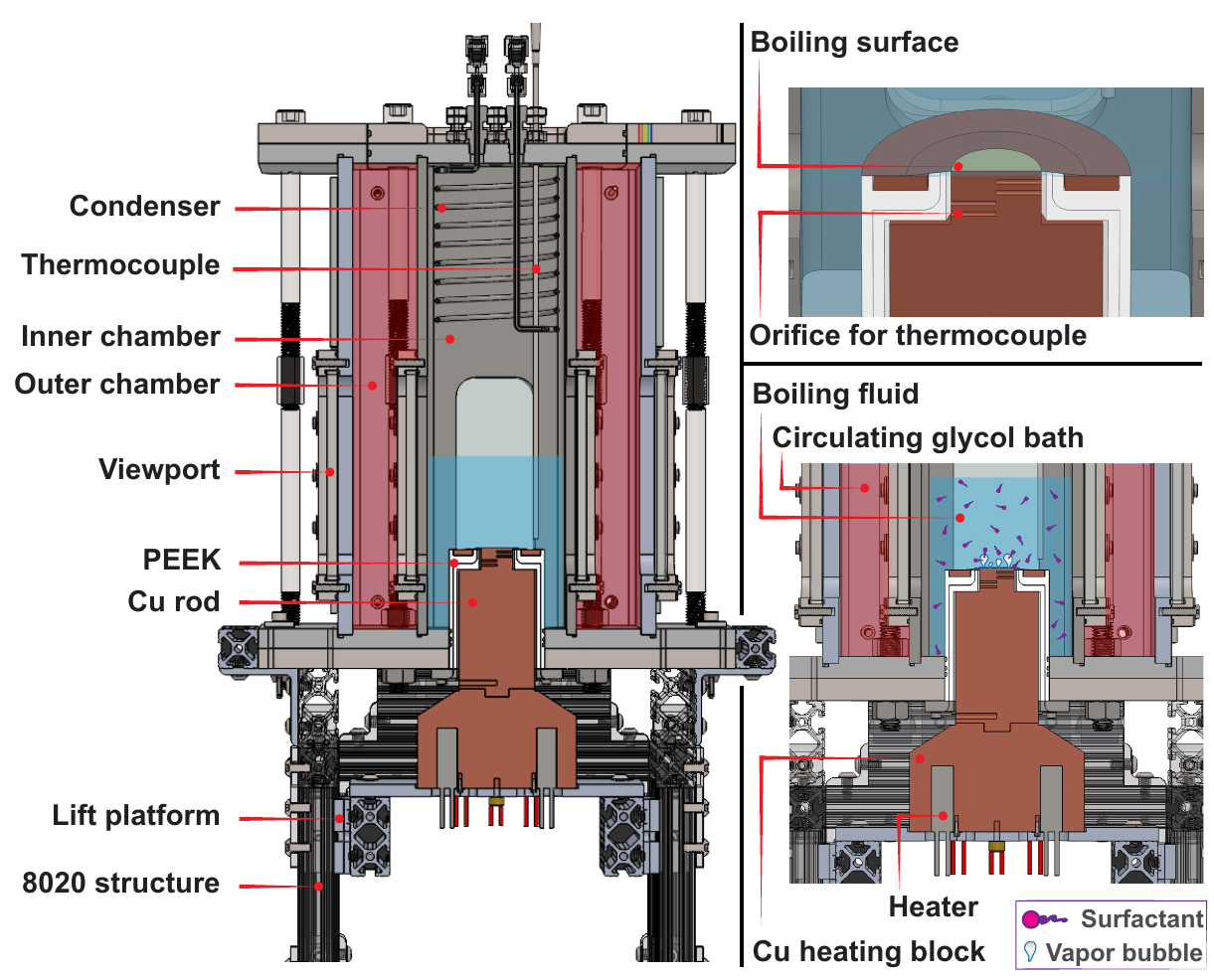}
    \caption{Schematic illustrating the boiling setup. In the outer chamber, there is a circulating ethylene glycol bath precisely controlled by a chiller unit that sets the bath at a temperature ($\SIrange{105}{110}{\celsius}$) slightly higher than the boiling point of the boiling fluid. This allows us to maintain $\SI{750}{\milli\liter}$ of boiling fluid at around $\SI{98}{\celsius}$ (saturation conditions in Las Vegas, NV, USA) throughout all the boiling experiments. Within the internal chamber, any vapor that accumulates is condensed by a stainless steel condenser, effectively preventing vapor from escaping into the surrounding atmosphere. Once the boiler reaches saturation conditions, we permit the boiling fluid to degas for approximately $\SI{30}{\minute}$ before commencing boiling experiments with deionized (DI) water. A series of DI water trials is conducted prior to each surfactant concentration assessment to confirm the continuous consistency of our boiling experiments. Once the boiling surface reaches CHF, we disengage the heating block from the copper rod by manually displacing the lift platform. The boiling surface cools by natural convection for about 30 minutes, and then we proceed with the next boiling experiment.}
    \label{BoilerSchematic}
\end{figure*}

\clearpage
\begin{figure*}[t]
    \centering
    \includegraphics[width=\textwidth]{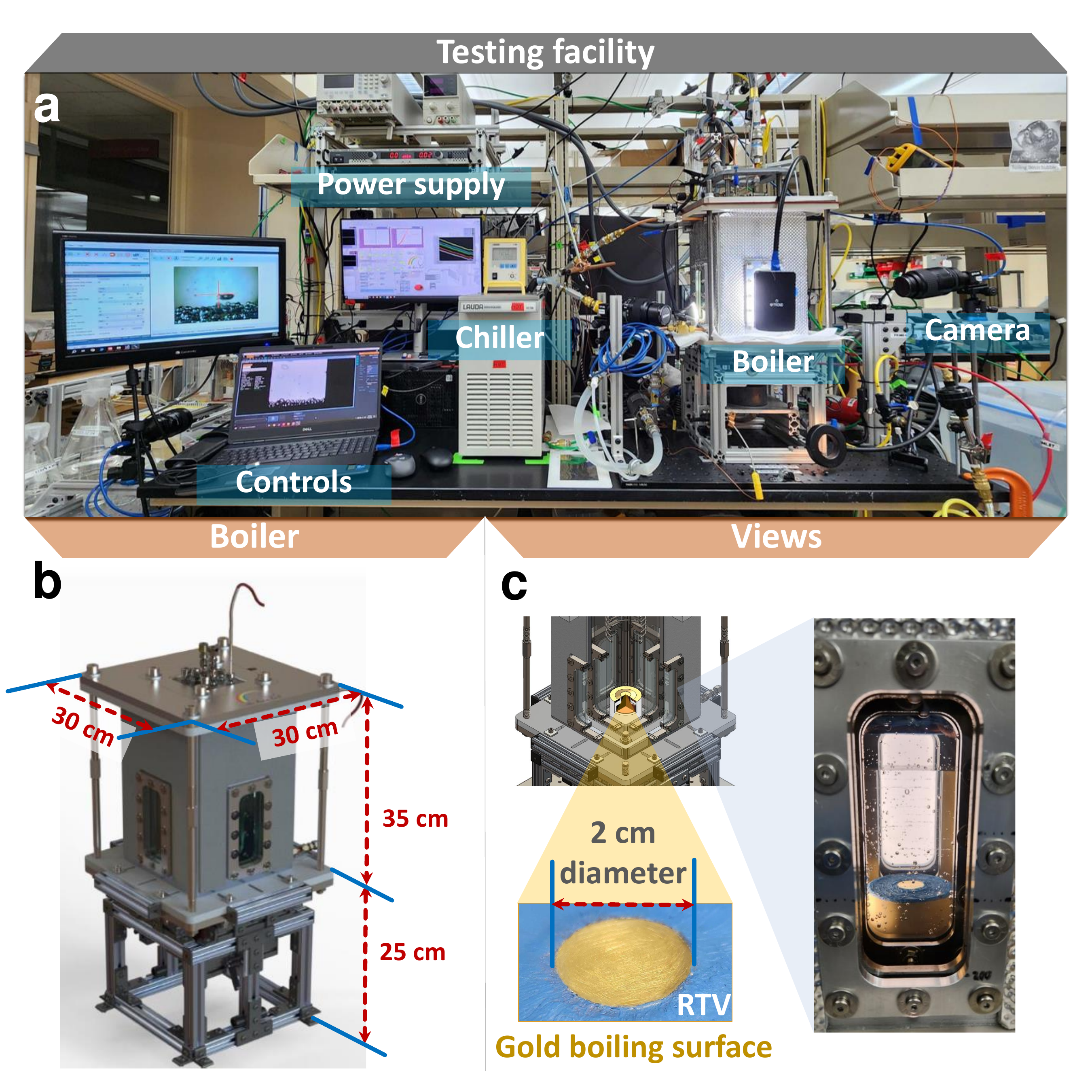}
    \caption{The testing facility in (a) shows the main equipment used to conduct boiling experiments and operate the boiler. The power supplied to eight cartridge heaters located in the copper heating block is controlled with a dedicated virtual instrument (VI) using LabVIEW. This VI reads the voltage signals from four thermocouples that are situated close to the boiling surface and connected to a data acquisition unit (MCC USB 8-channels). These signals are immediately transmitted to a code we wrote in Wolfram Mathematica to extrapolate the boiling surface temperature and calculate heat flux. This information is then displayed live on a monitor in the LabVIEW software. The boiler is illuminated with several LED lamps, and bubble visualization is obtained through two cameras: a polarized FLIR Blackfly-S camera on the side and a high-speed Photron AX-200 camera on the back of the boiler. Rendered view of (b) the boiler with its primary dimensions. The section view in (c) shows the exposed surface of the gold boiling sample. We sanded the boiling surface with 320-grit sandpaper to promote nucleation sites. The gold boiling surface was soldered on top of a copper foil, and this copper foil was then soldered onto a copper rod. We applied a thin layer of blue RTV silicone adhesive sealant on top of all the foil exposed to the working fluid, except for the boiling region, which has a $\SI{2}{\centi\meter}$ diameter for boiling experiments.}
    \label{BoilerEquipment}
\end{figure*}

\clearpage
\begin{figure*}[t]
    \centering
    \includegraphics[width=\textwidth]{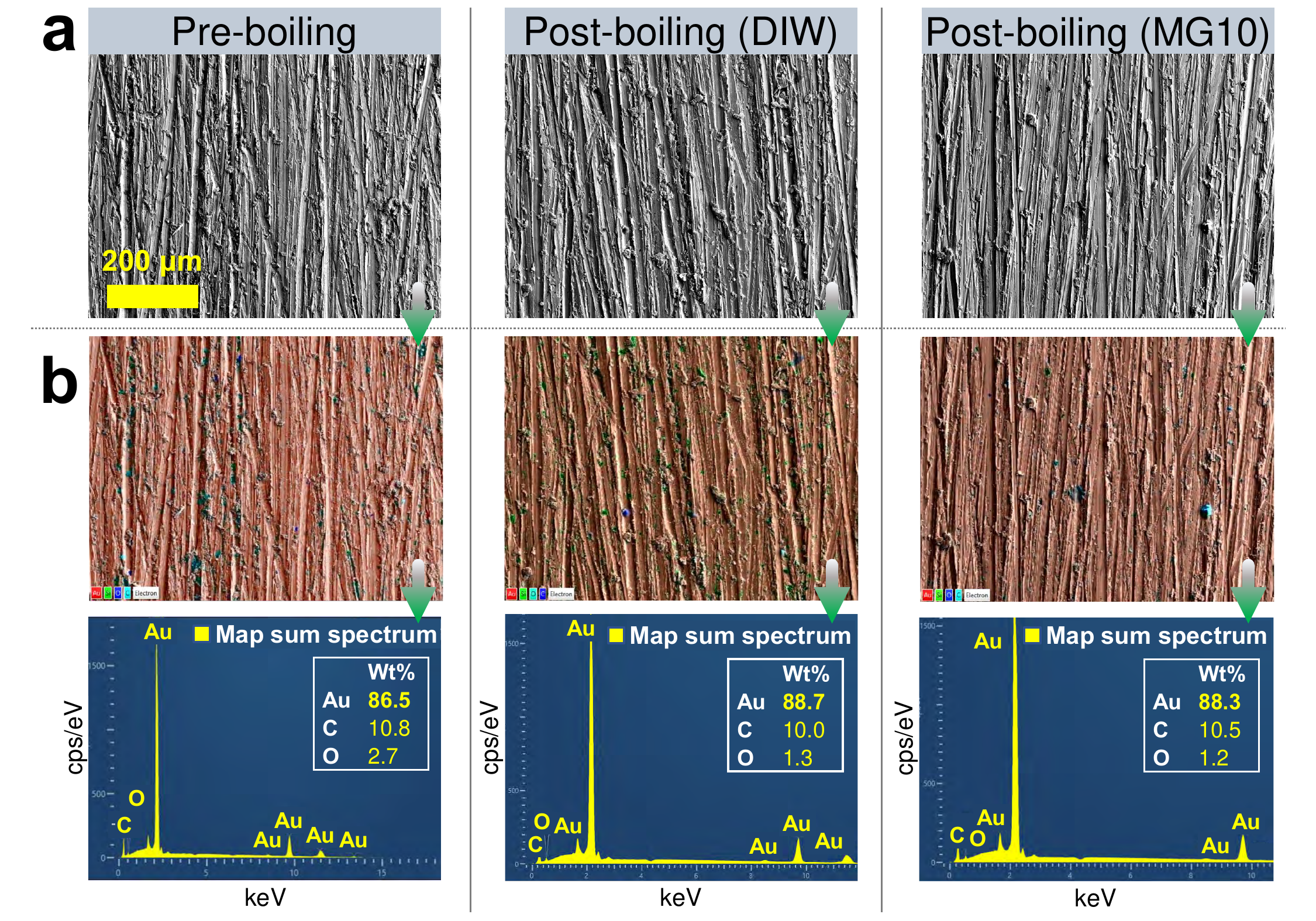}
    \caption{We conducted field emission scanning electron microscopy (FESEM) on small pieces extracted from the gold boiling surface. In (a), we can observe images of the gold sample in its pristine state before exposure to boiling water (pre-boiling), after being submerged in boiling water (post-boiling (DIW)), and after being exposed to the boiling aqueous solution with surfactant additives of MEGA-10 (post-boiling (MG10). The surfactant MG10 was the final surfactant additive we employed. Therefore, we conducted this FESEM analysis on the gold surface exposed to this last surfactant additive. The FESEM analysis indicates that there are no discernible alterations to the surface structure, suggesting that there was either no oxidation or only minimal effects on the pure gold sample due to the boiling fluid treatment. Turning to section (b), we present the results of energy-dispersive X-ray spectroscopy (EDS) for the three samples. These results have been spatially mapped across the entire area visible in the FESEM images from section (a) and quantified as counts per second per electron-volt (cps/eV) plotted against kilo-electron-volt (keV). These findings illustrate a high concentration of gold in the chemical composition (close to $\SI{90}{\percent}$, along with some traces of carbon and oxygen that were adsorbed onto the sample's surface). 
    }
    \label{SEMImages}
\end{figure*}

\clearpage
\begin{figure*}[t]
    \centering    \includegraphics[width=\linewidth, angle=270]{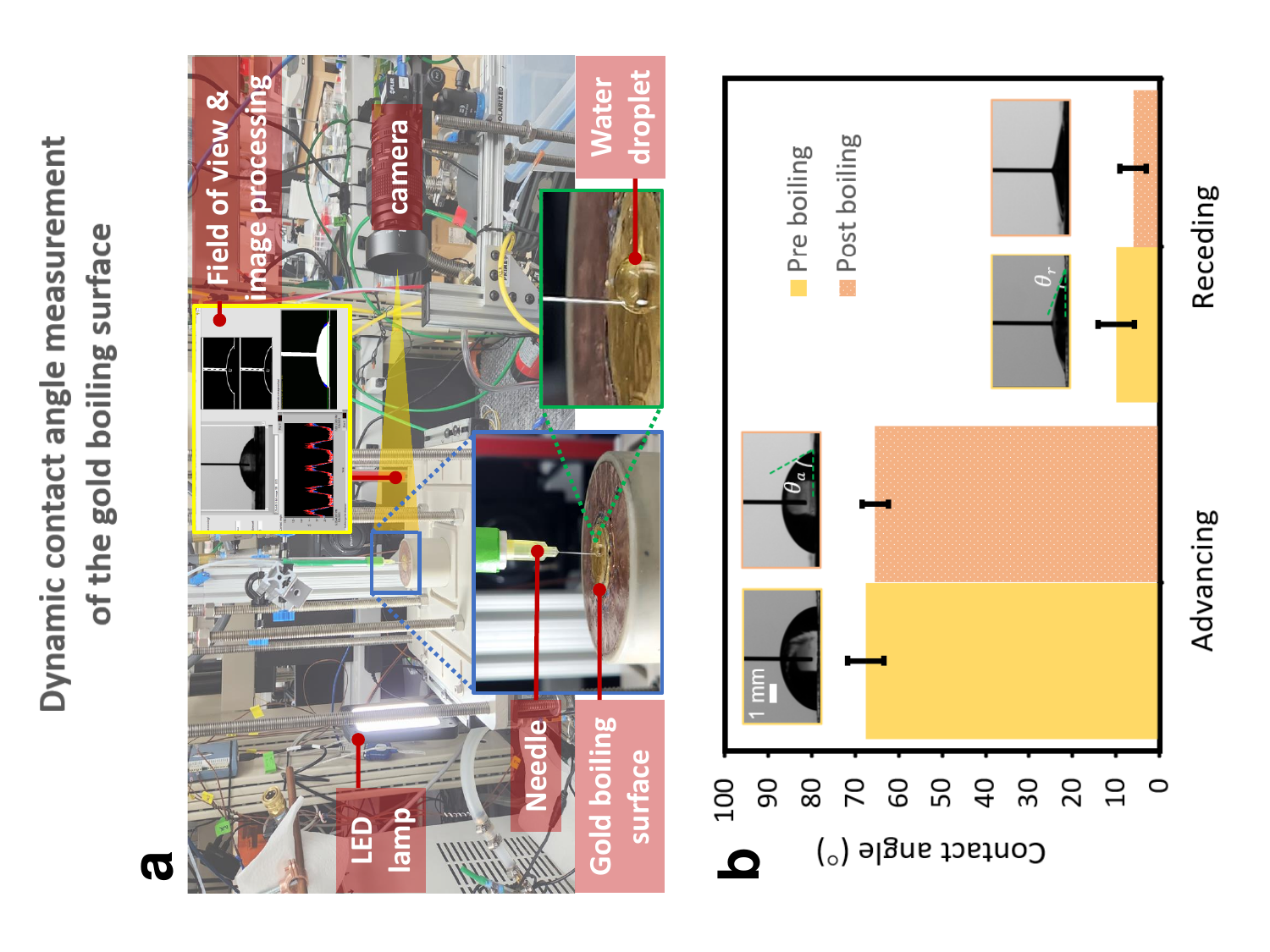}
    \caption{Dynamic contact angle measurements of our pristine gold boiling surface for (a) a fresh new sample, prior to boiling experiments (pre-boiling) and after being exposed to boiling liquid (post-boiling). In (b), we display our in-situ goniometer (with the boiling setup dissembled). We precisely control our goniometer using a PID algorithm in LabVIEW and a code written in Wolfram Mathematica. At the time we took the picture in (a), we already removed the RTV silicone compound that is generally applied on the exposed copper surface, which is a disk surrounding the gold foil and is not part of the boiling surface. For more details about the RTV on the mentioned copper disk, see Fig.~\ref{BoilerEquipment}.}
    \label{DIWGoldCA}
\end{figure*}

\clearpage
\begin{figure*}[t]
    \centering
    \includegraphics[width=1\textwidth]{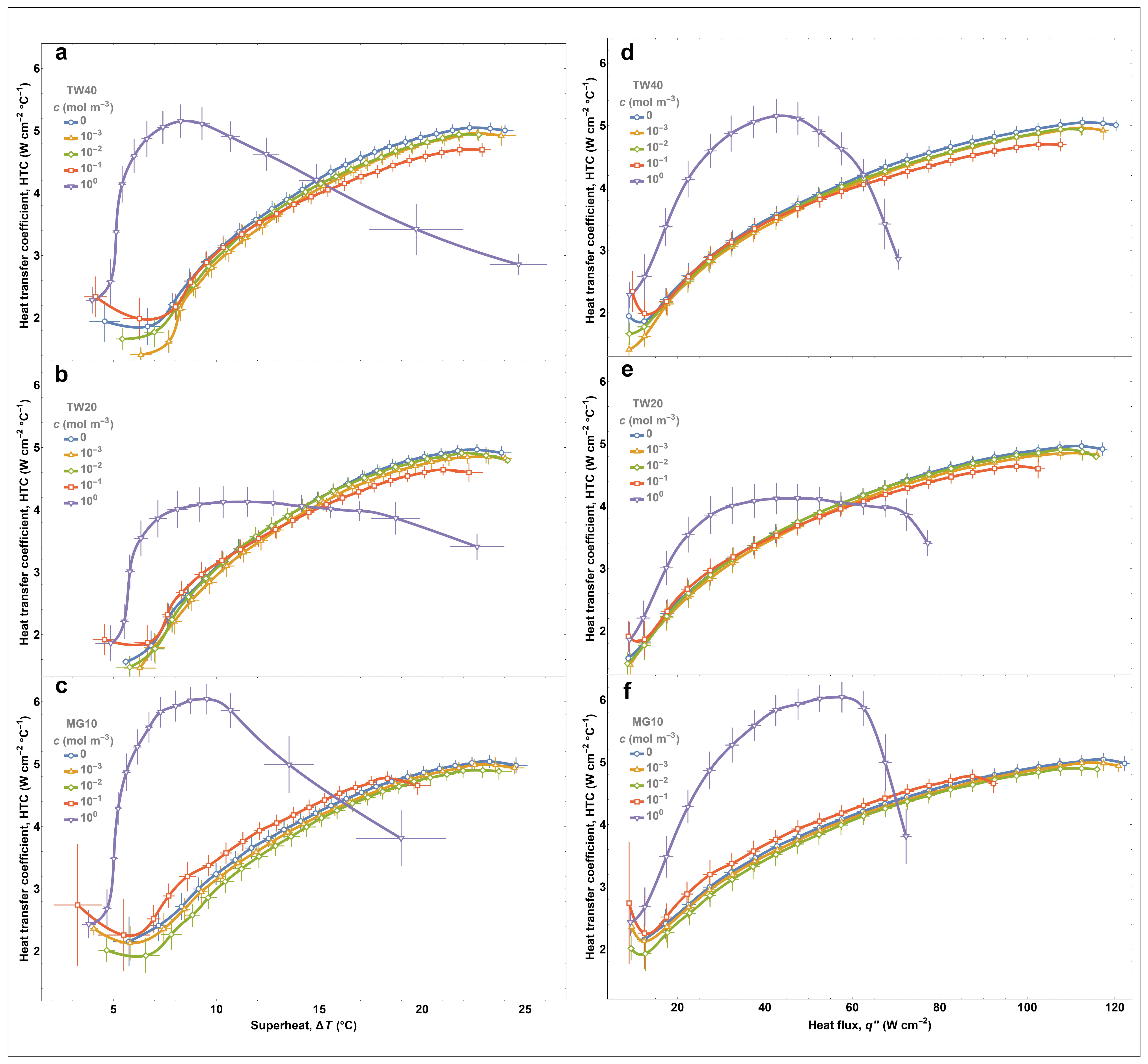}
    \caption{We depict in (a-c) the HTC in relation to superheat for the three surfactants (TW40, TW20, and MG10). Similarly, we show the HTC against heat flux in (d-f) to illustrate the heat flux range where the maximum HTC occurs in our experiments. The HTCs peak in the heat flux range of $\SIrange{10}{50}{\watt\per\centi\meter\squared}$ for the three surfactants at different molar concentrations. Compared to DI water ($\SI{0}{\mole\per\meter\cubed}$), we observe minimal HTC changes at lower concentration values ($\SI{0.001}{\mole\per\meter\cubed}$, $\SI{0.01}{\mole\per\meter\cubed}$, and $\SI{0.1}{\mole\per\meter\cubed}$ for surfactant TW40 in (a). However, at a concentration of $\SI{1}{\mole\per\meter\cubed}$, TW40 reaches a peak in HTC with a low superheat range of $\SIrange{5}{10}{\celsius}$. This trend remains consistent in (b) and (c) for surfactants TW20 and MG10, with low superheats aligning with HTC peaks. Among these three nonionic surfactants, at a molar concentration of $\SI{1}{\mole\per\meter\cubed}$, TW20 exhibits the lowest HTC, around $\SI{4}{\watt\per\centi\meter\squared\per\celsius}$. TW40 shows a maximum HTC close to $\SI{5}{\watt\per\centi\meter\squared\per\celsius}$, while MG10 achieves the highest HTC at about $\SI{6}{\watt\per\centi\meter\squared\per\celsius}$. These peak HTC values are observed within a heat flux range of $\SIrange{10}{50}{\watt\per\centi\meter\squared}$ in (d-f). Consequently, any comparison of boiling performance between these surfactants in this study involves averaging the HTC within the heat flux range of $\SIrange{10}{50}{\watt\per\centi\meter\squared}$.
}
    \label{HTCPlots}
\end{figure*}

\clearpage
\begin{figure*}[ht]
    \centering
    \includegraphics[width=0.9\textwidth]{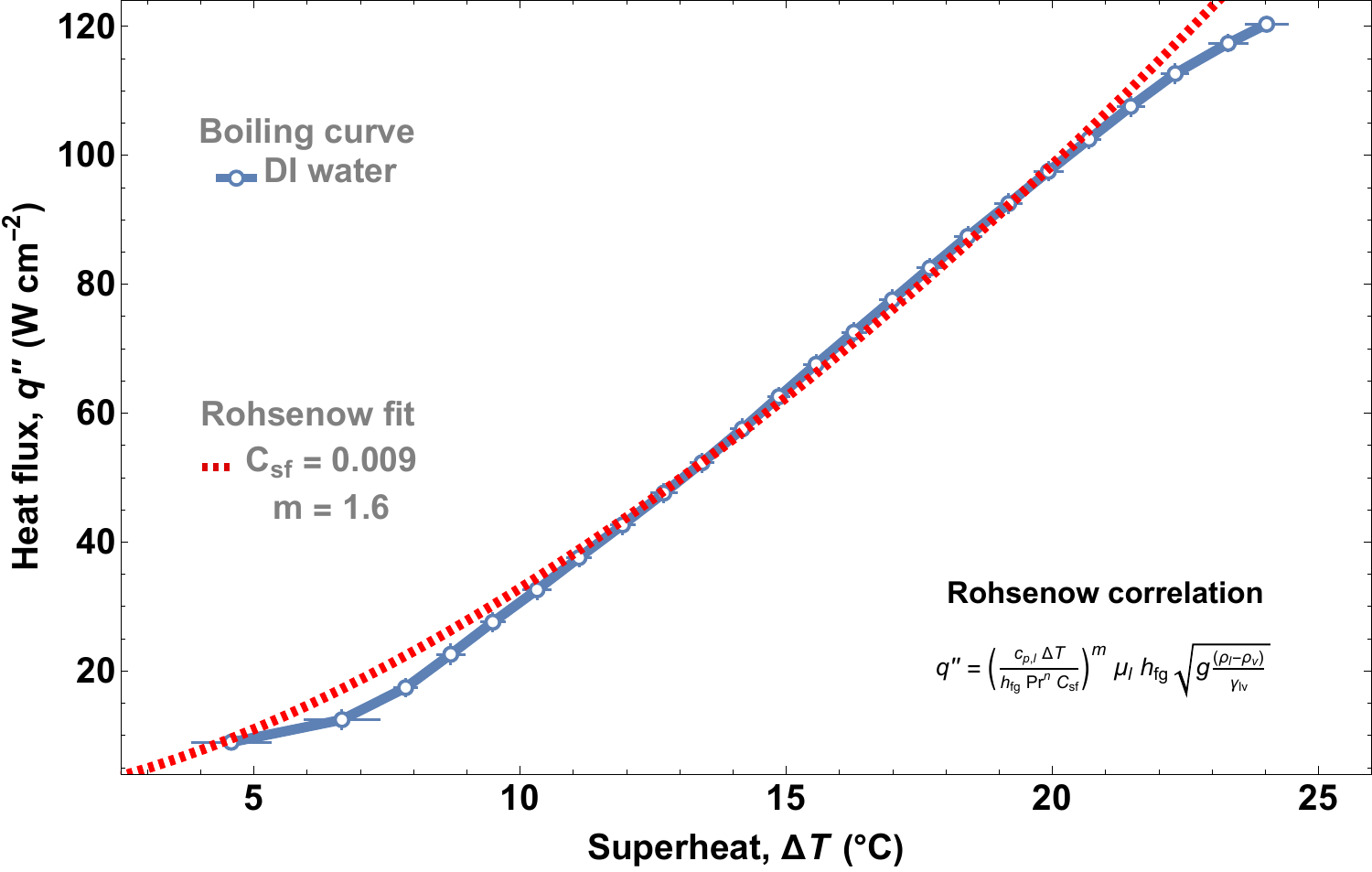}
    \caption{The averaged boiling curve for deionized (DI) water (reference boiling curve of TW40 without surfactants ($c = \SI{0}{\mole\per\meter\cubed}$) from Fig.~\ref{BoilingCurves}a) aligns well with the Rohsenow correlation (equation in this figure) featuring a fitted surface fluid factor, $C_\text{sf}$, equal to 0.009, and an exponent, $m$, equal to 1.6.}
    \label{RohsenowFit}
\end{figure*}

\clearpage
\begin{figure*}[ht]
    \centering
    \includegraphics[width=\textwidth]{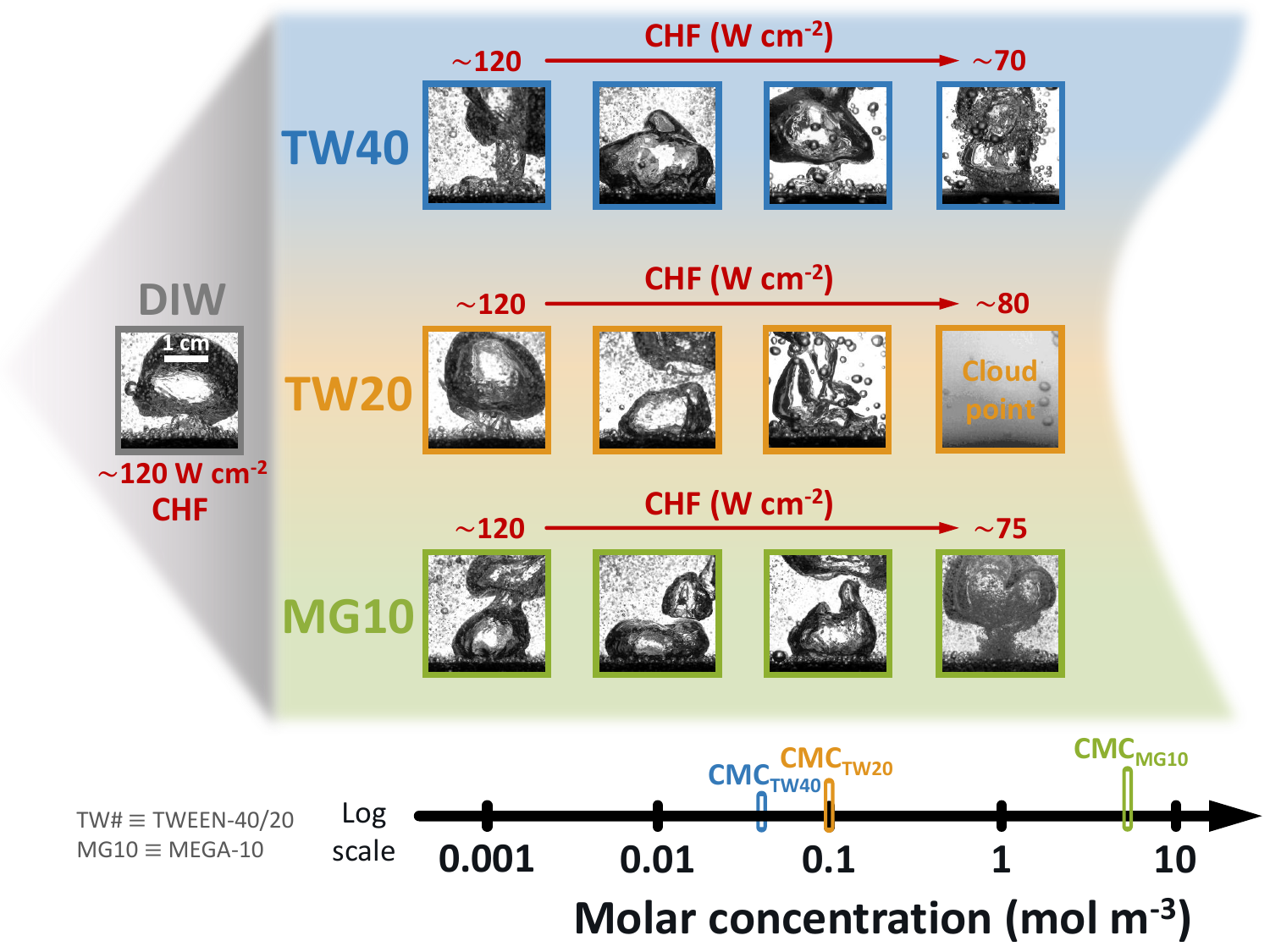}
    \caption{High-speed bubble visualization of the three tested surfactants (TW40, TW20, and MG10) at different molar concentrations close to their respective CHFs did not reveal significant changes in boiling behavior at these high heat fluxes. Our high-speed camera failed to capture bubble formation on the boiling surface of TW20 at $\SI{1}{\mole\per\meter\cubed}$, primarily due to the exacerbated cloud point. Positioned on the back side of our boiler, our high-speed camera captured these bubble snapshots during the same experiments that we illustrated using a secondary polarized camera, which captured the bubbling events at lower heat flux (see Fig.~\ref{BubblingBehavior}).}
    \label{CHFBubblingBehavior}
\end{figure*}

\clearpage
\begin{figure*}[t]
    \centering
    \includegraphics[width=\textwidth]{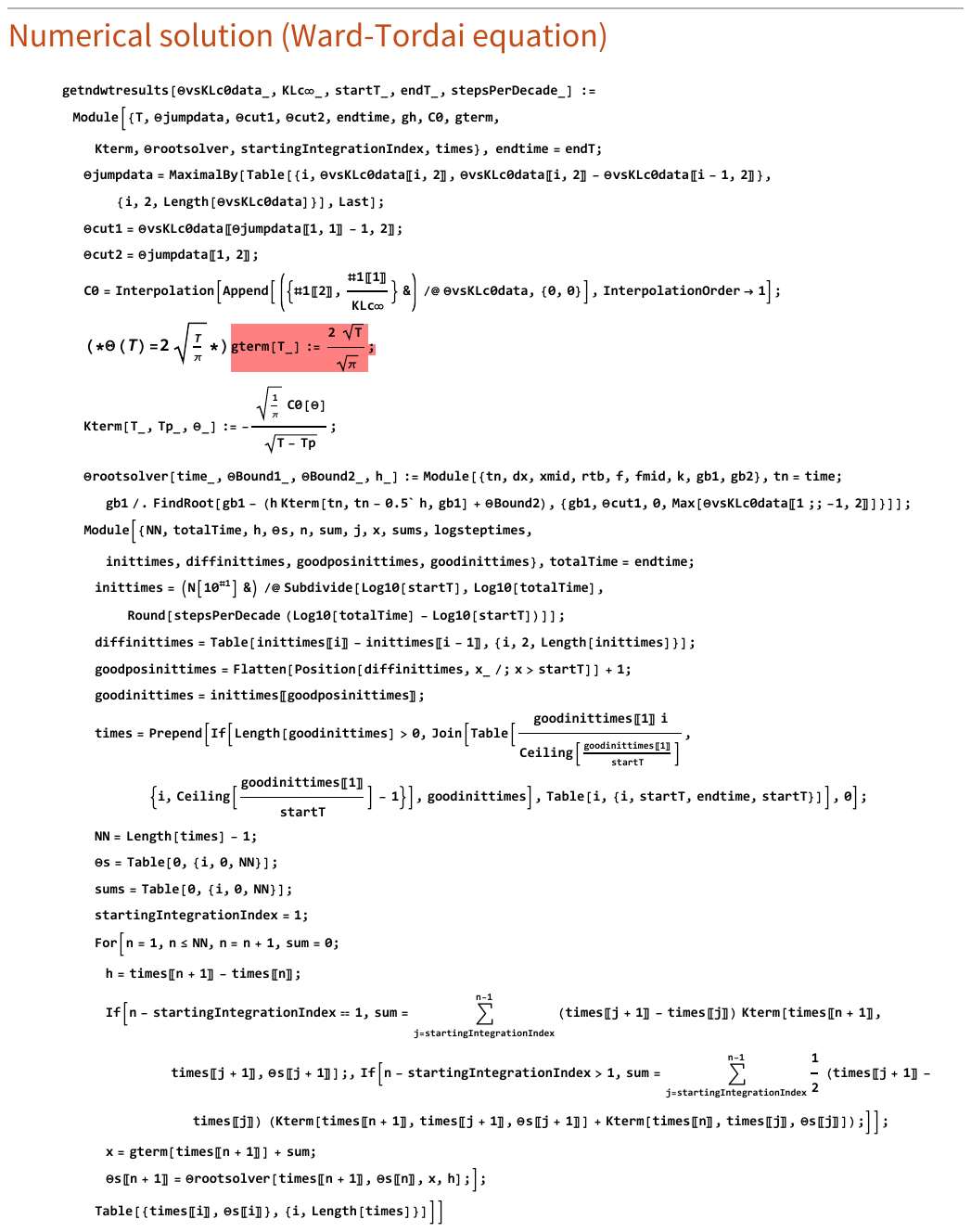}
    \caption{Code used to numerically solved the Ward-Tordai equation~\cite{Ward1946} using an algorithm by Li et al.~\cite{Li2010}.}
    \label{Code1}
\end{figure*}

\clearpage
\begin{figure*}[ht]
    \centering
    \includegraphics[width=\textwidth]{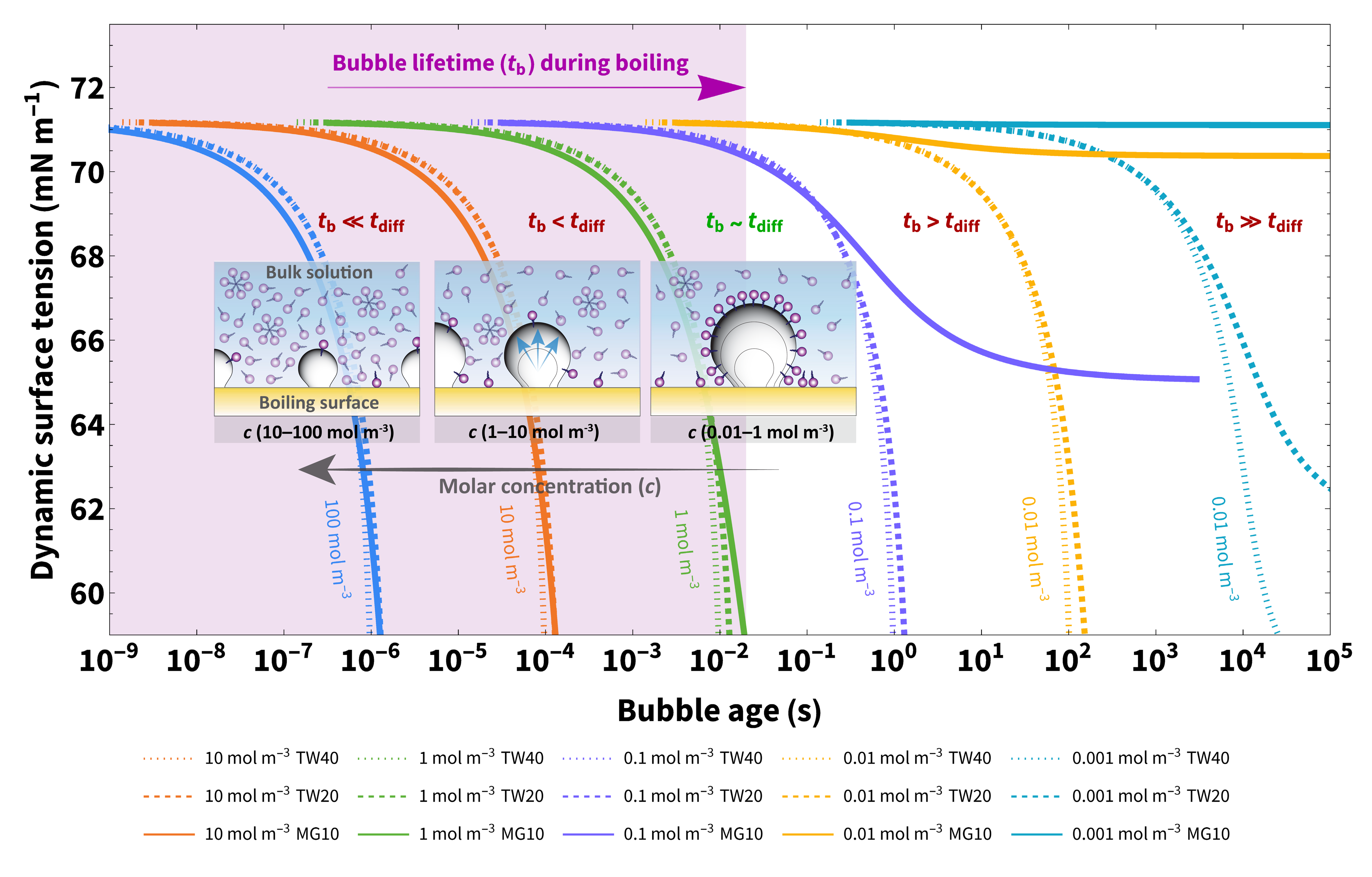}
    \caption{The modeled dynamic surface tension results indicate a similar decrease in surface tension at short bubble age timescales below the estimated boiling bubble lifetime ($\sim\SI{20}{\milli\second}$) for all surfactants at the same molar concentrations. As the concentration increases, the timescale of surface tension decreases. At a concentration around $\sim\SI{1}{\mole\per\meter\cubed}$, all surfactants exhibit a similar decrease within the boiling bubble lifetime. The dynamic surface tension data was obtained by numerically solving the Ward-Tordai equation (see Eq.~\ref{eqn:wardtordaimodified}), plugging data from Table~\ref{TAB1} in our dedicated algorithm developed in Mathematica (see Appendix A, Fig.~\ref{Code2}).}
    \label{DynamicSurfaceTension}
\end{figure*}

\clearpage
\begin{figure*}[t]
    \centering
    \includegraphics[width=\textwidth]{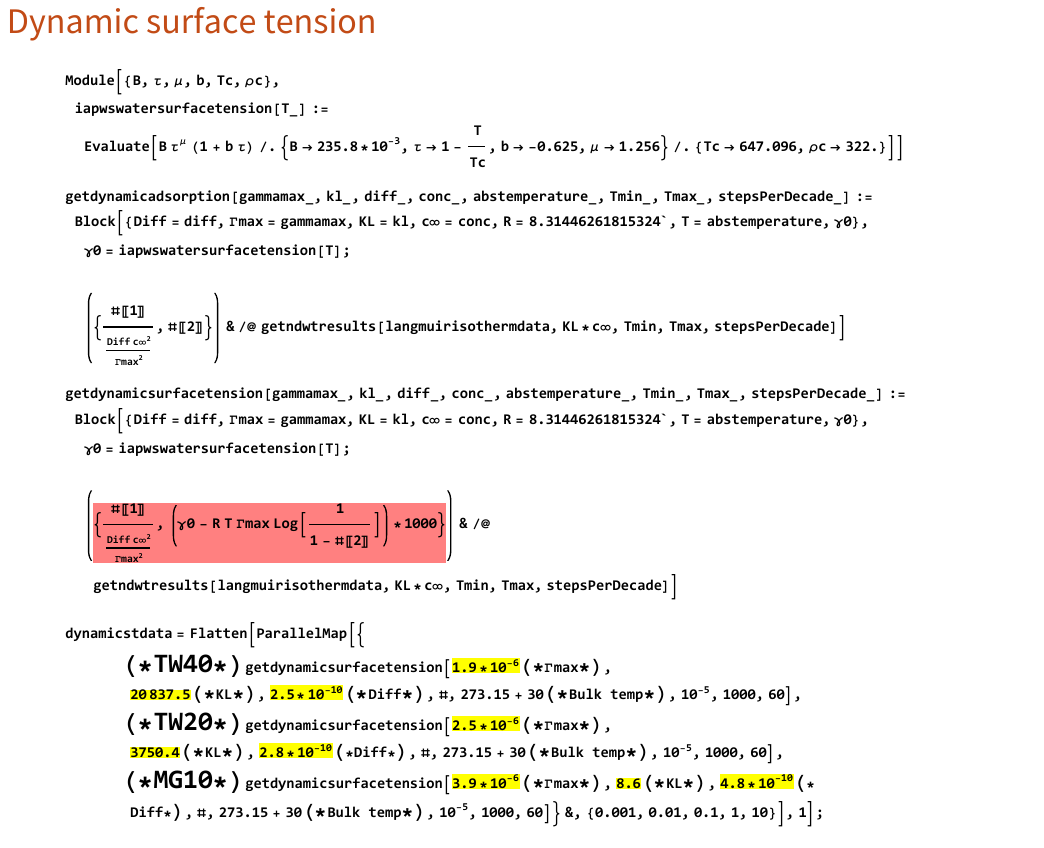}
    \caption{Code used to model dynamic surface tension (see Appendix A, Fig.~\ref{DynamicSurfaceTension}).}
    \label{Code2}
\end{figure*}

\clearpage

\section{Materials, methods, and results of Boiler 2}
\label{APPENDIXB}
In this appendix, we detail the experimental setup, measurement protocol, and materials employed with the second boiler, which features a gold-sputtered copper surface. This boiler is the property of our co-authors, Matic Može, Armin Hadžić, and Iztok Golobič. The experiments were conducted in Ljubljana, Slovenia.

\subsection{Experimental setup}
\begin{figure*}[t]
    \centering    \includegraphics[width=\linewidth]{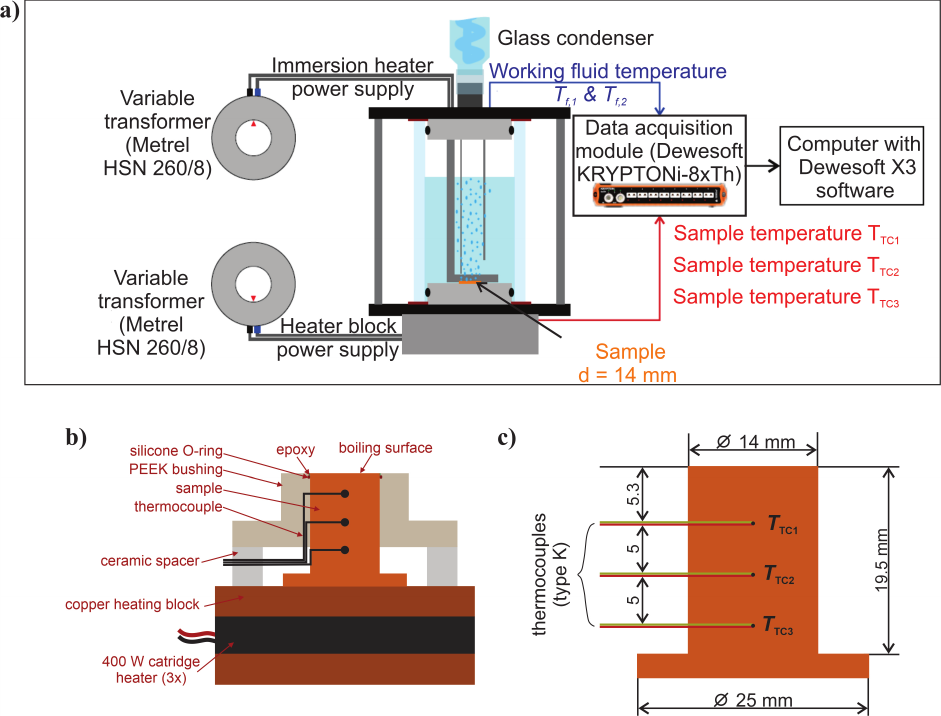}
    \caption{a) Experimental setup, b) heating block and sample assembly, and c) sample dimensions including thermocouple locations.}
    \label{MaticBoiler}
\end{figure*}

We conducted the assessment of pool boiling performance using a customized experimental apparatus (refer to Appendix B, Fig.\ref{MaticBoiler}a). Specifically, we drilled three holes, each measuring $\SI{7}{\milli\meter}$ in length and $\SI{0.8}{\milli\meter}$ in diameter, evenly spaced at $\SI{5}{\milli\meter}$ intervals along the lateral surface of the sample (refer to Appendix B, Fig.~\ref{MaticBoiler}b and Fig.~\ref{MaticBoiler}c). Thin thermocouples of the K-type were embedded in these holes. We affixed the sample to a copper heating block, which we subsequently inserted into the boiling chamber through the lower stainless-steel flange. We used a PEEK bushing, a ring of flexible epoxy glue, and a silicone O-ring to provide sealing, limit heat loss, and prevent parasitic boiling (refer to Appendix B, Fig.~\ref{MaticBoiler}b). The boiling chamber itself was fabricated using a glass cylinder with an inner diameter of $\SI{60}{\milli\meter}$, flanked by two stainless steel flanges. We filled the boiling chamber with approximately $\SI{200}{\milli\liter}$ of working fluid. The heat required for boiling was generated by cartridge heaters positioned within the heating block, which were also controlled by a variable transformer. All measurements were performed at atmospheric pressure, and we confirmed its stability by a stable saturation temperature. The vapor produced during the measurements was fed into the water-cooled glass condenser and returned to the boiling chamber.

Two other K-type thermocouples were immersed in the boiling chamber at different heights to measure the temperature of the working fluid. We used a KRYPTONi‐8xTH DAQ device to collect all temperature signals as raw voltages. Data from the DAQ device were acquired using Dewesoft X3 software at a frequency of $\SI{60}{\hertz}$. A detailed description of the experimental setup used for boiling evaluation was presented in our previous research~\cite{Hadzic2022, MMoze2022}. Moreover, the experimentation involved acquiring five temperature data points during the boiling test. We derived Instantaneous values of pertinent heat transfer parameters through real-time computations using DewesoftX software both during data collection and subsequent analysis using MathWorks MATLAB. We recorded the data at a sampling rate of $\SI{10}{\hertz}$, and $\SI{1}{\hertz}$ averaged values were computed. Heat flux determination was accomplished by leveraging the spatial temperature gradient within the sample, assuming one-dimensional conduction in accordance with Fourier's law of heat conduction. This calculation incorporated the temperature-dependent thermal conductivity of the sample, evaluated at its average temperature.

We extrapolated the surface temperature of the sample by using the temperature of the thermocouple closest to the boiling surface and the calculated heat flux. Concurrently, the surface superheat, denoting the temperature difference between the sample's surface and the average temperature of the water within the boiling vessel, was established. Finally, we calculated the heat transfer coefficient by dividing the heat flux by the surface superheat to obtain a metric describing the boiling performance and heat transfer intensity. A detailed description of data reduction and measurement uncertainty calculations including all relevant equations are provided in our previous works~\cite{MMoze2022}. 

\subsection{Measurement protocol}
We employed a uniform measurement protocol to assess the heat transfer efficiency of various water-surfactant solutions. Subsequent to sputter-gold plating the sample's surface, we affixed the sample to the heater assembly, with the edges sealed using epoxy to prevent parasitic nucleation. This assembly was then affixed through the lower flange of the boiling vessel, and the identical sample was consistently employed for testing all water-surfactant solutions, including TWEEN-20, TWEEN-40, and MEGA-10. To establish baseline results for surface boiling performance, the surface was initially evaluated using twice-distilled water prior to testing each of the aforementioned water-surfactant solutions. Furthermore, each water-surfactant solution, starting from the lowest ($\SI{0.001}{\mole\per\meter\cubed}$) concentration and progressing to the highest ($\SI{1}{\mole\per\meter\cubed}$), underwent testing. All tested solutions, along with twice-distilled water, were subjected to saturation and degassing through vigorous boiling for a minimum of $\SI{45}{\minute}$ before conducting the measurements. During this period, we applied a heat flux of approximately $\SI{400}{\kilo\watt\per\meter\squared}$ to the sample to induce nucleate boiling and facilitate the removal of potentially entrapped air on the surface. Twice-distilled water along with each concentration of water-surfactant solution was subjected to three repeated measurements to ensure measurement stability.

Prior to initiating an experimental run, both the sample and the water were cooled to $\SI{90}{\celsius}$ to condense any trapped vapor. Subsequently, the temperature of the water-surfactant solution was raised to saturation, while we gradually increased the heating power applied to the cartridge heaters beneath the sample. Throughout the measurement process, we incrementally raised the heating power at a slow rate of approximately $\SI{0.2}{\kilo\watt\per\meter\squared\per\second}$ within the natural convection regime and up to $\SI{2}{\kilo\watt\per\meter\squared\per\second}$ within the nucleate boiling regime. This approach allowed for a more rapid assessment of the boiling curve and mitigated the potential influence of prolonged boiling and exposure to water on the boiling surface. We empirically and numerically verified this methodology to be a robust approximation for steady-state measurements. We conducted the measurements until we reached CHF, at which point the heaters were deactivated, and the sample was permitted to cool below the water's saturation temperature. Subsequently, the measurement process was reiterated, with a total of three runs carried out for each tested concentration of the water-surfactant solution and also for twice distilled water.

To ensure the accuracy of subsequent measurements with higher concentration solutions, we performed a thorough cleaning on both the boiling chamber and sample using deionized water. This served to eliminate any remnants from previous experiments involving water-surfactant solutions of lower concentration. Furthermore, following experiments with one type of water-surfactant solution, a thorough cleaning utilizing deionized water and isopropanol was implemented to remove surfactant residues before conducting experiments with the next type of water-surfactant solution.

Used surfactants:

TWEEN 20: (Tween 20, Carl Roth).

TWEEN 40: (Tween 40, Carl Roth).

MEGA-10: (MEGA-10, Avanti POLAR LIPIDS2, INC.).

\subsection{Boiling curves of Boiler 2 (gold-sputtered copper surface)}

\begin{figure*}[ht]
    \centering    \includegraphics[width=0.7\linewidth]{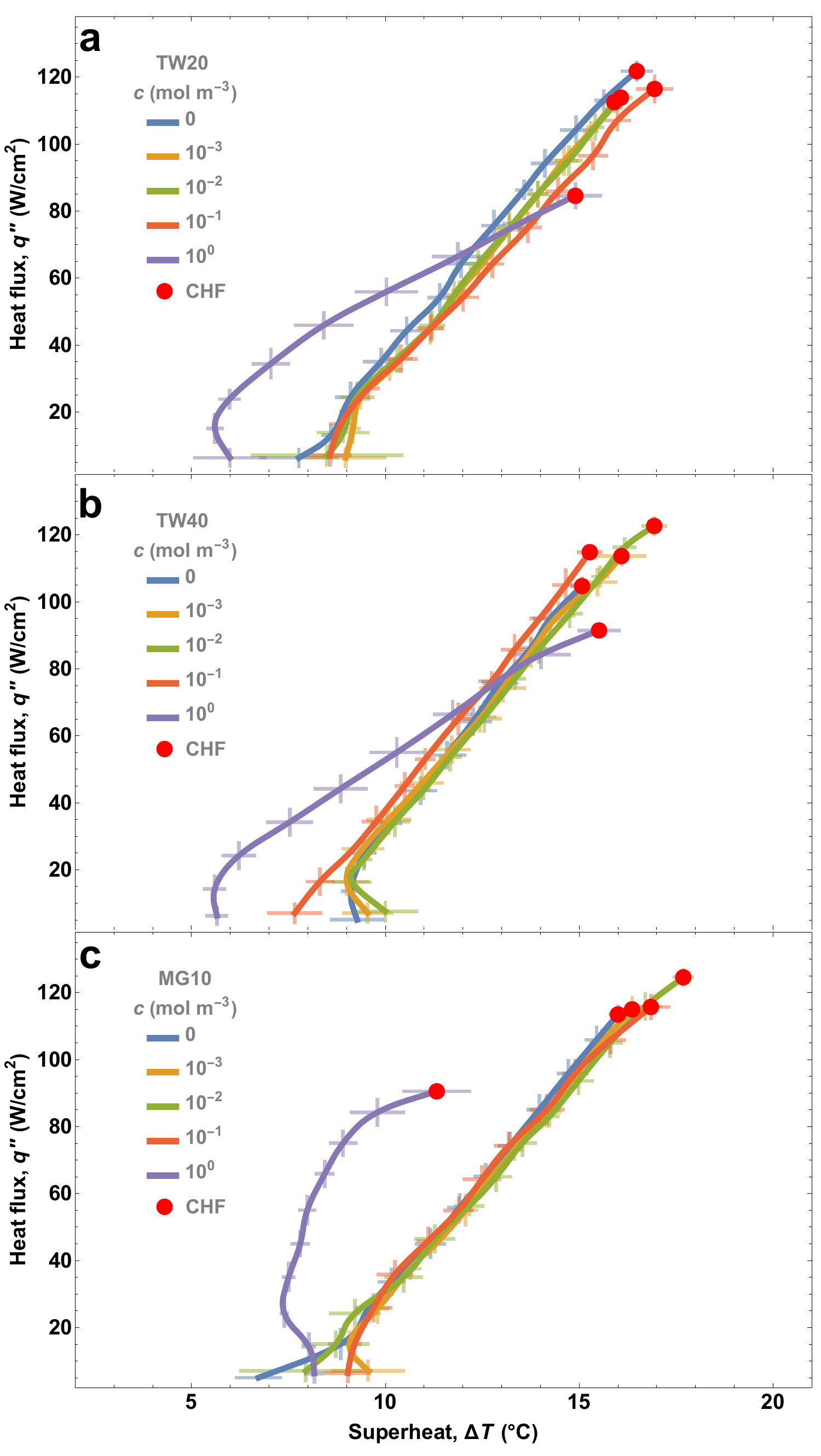}
    \caption{Experimental boiling results of three surfactants, TWEEN-40 (TW40), TWEEN-20 (TW20), and MEGA-10 (MG10) show the surfactant-induced effects in the boiling performance (boiling curves shifting to the left) as we increase the molar concentration of the bulk solution. These experiments were obtained with a boiling surface of different size ($\SI{14}{\milli\meter}$ diameter), material (gold-sputtered copper), and boiling conditions. However, the results are aligned with the findings of this work (see Fig.~\ref{BoilingCurves}) where minimal changes in the CHF are observed in the low concentration range $\SIrange{0.001}{0.1}{\mol\per\meter\cubed}$. The most substantial change in CHF occurs at a molar concentration of $\SI{1}{\mol\per\meter\cubed}$, a concentration shown in our study that makes the bubble timescale ($t_\text{b}$) near the timescale of diffusion ($t_\text{diff}$).}
    \label{BCsMatic}
\end{figure*}






\end{document}